\documentclass[11pt,a4paper]{article}
\usepackage{jheppub_kim}
\usepackage{epsfig}
\usepackage{amsmath}
\usepackage{graphicx}
 \usepackage{graphics}
 \usepackage[scriptsize,nooneline,hang]{caption}
\usepackage[hang,nooneline,scriptsize]{subfigure}
 
\begin{document}

\title{Phase transition of a charged AdS black hole with a global monopole through geometrical thermodynamics}
 
\author[a]{Saheb Soroushfar, }
\affiliation[a]{Faculty of Technology and Mining, Yasouj University, Choram 75761-59836, Iran}
\author[b,c]{Sudhaker Upadhyay }
\affiliation[b]{Department of Physics, K.L.S. College, Nawada-805110 (A Constituent Unit of Magadh University, Bodh-Gaya), India }
\affiliation[c]{Visiting Associate, Inter-University Centre for Astronomy and Astrophysics (IUCAA) Pune, Maharashtra-411007}
\emailAdd{soroush@yu.ac.ir}
\emailAdd{sudhakerupadhyay@gmail.com}

\abstract
{In order to study the phase transition through thermodynamic geometry, we consider   the  charged AdS black hole with global monopole. 
We first introduce   thermodynamics of charged AdS black hole with global monopole 
by discussing the dependence of Hawking temperature, specific heat and $P-v$ curve 
on horizon radius and monopole parameter. By implementing various thermodynamic geometry methods, for instance, Weinhold, Ruppiner, Quevedo and HPEM formulations, we derive corresponding scalar curvatures  for charged AdS black hole with a global monopole. 
Here, we observe that, in contrast to Weinhold and Ruppeiner methods,  HPEM and Quevedo formulations provide more information about the phase transition of the charged AdS black hole with a global monopole.  }


\maketitle

\section{Introduction} 	  
The first notion of relationship between black holes and thermodynamic systems came 
when Bekenstein  found that thermodynamic entropy is closely analogous to the area
of a black hole event horizon \cite{bak}. The cornerstone of this relationship is black hole thermodynamics (mechanics), which states that   black hole also follows  analogous  laws to the
ordinary thermodynamics. This is justified by Hawking by stating that black holes can radiate
\cite{haw}.  
The black holes can be categorized in two forms,
larger ones with positive specific heat which are locally stable, and smaller ones with negative
specific heat which are unstable. The phase transition between  black holes and
radiation at the transition temperature is known as Hawking Page phase transition \cite{haw1,page}.
 The importance of AdS black holes lies to the fact that   the thermodynamically
stable black holes exist only in AdS space.

Monopoles are  nothing but the defects like cosmic strings and domain walls which were originated during the cooling phase of the early universe \cite{mon,mon1}.  It is found that a global monopole charge accompanies spontaneous breaking of
global $O(3)$ symmetry into $U(1)$ in phase transitions in the Universe \cite{Barriola:1989hx}.  Here, the static black hole solution with a global monopole was  obtained 
 with different   topological structure than the
Schwarzschild black hole solution. 
In literature, there are various discussions on the black hole solution with
monopole  \cite{mon2,mon3,mon4,mon5,mon6,mon7,mon8}. Here, it is confirmed that the presence of the global monopole in the black hole solution  plays significant role.

Meanwhile,  there exist several ways   to implement differential geometric concepts in thermodynamics of black holes \cite{sud,sud1,sud2}. 
One of them is to formulate the concept of thermodynamic
length, Ruppeiner  metric \cite{Ruppeiner}, a conformally equivalent to Weinhold's
metric \cite{Weinhold}. Here, it is found that the phase space and the metric structures suggested that Weinhold's and Ruppeiner's  metrics are not invariant under Legendre transformations \cite{sal,mru}. Then, an attempt was made by  Quevedo \cite{Quevedo} to unifies the geometric properties of the phase space and the
space of equilibrium states.  
Moreover,  Hendi--Panahiyan--Eslam--Momennia (HPEM) metric  
is also studied, which builds a geometrical phase space by thermodynamical quantities \cite{Hendi:2015rja,Hendi:2015xya,EslamPanah:2018ums}. Our motivation here is to study the phase transition of a charged AdS black hole with a global monopole through geometrical thermodynamics.

In this regard, we first consider a  a charged AdS black hole solution with global monopole.
In tandem to the Hawking temperature, specific heat and electric potential, we discuss critical parameters for this black hole.
In fact, behavior of temperature with respect to horizon radius for different values of charge, AdS radius and monopole parameter are discussed. Here, we find that   the temperature first increases to a maximum point and then  decreases  with increasing values of horizon radius.  After a certain value  of horizon radius, temperature  only increases  with horizon radius.  
The behavior of heat capacity for different values of charge, AdS radius and monopole parameter   are also studied. Remarkably,  for certain values of these parameters, heat capacity has one zero point describing a physical limitation point. Also,  heat capacity has two divergence points, which  demonstrate phase transition critical points of a charged AdS black hole with a global monopole. 
We plot $P-v$ diagrams of the charged AdS black hole with a global monopole  in extended phase space for specific value of parameters.  
Under a critical temperature a critical behavior  occurs which distorted  with larger values of energy scale. Also, with larger values of this energy scale, in contrast to  critical temperature and critical pressure, critical volume increases.

We also analyse the geometric structure of Weinhold, Ruppiner, Quevedo and HPEM
formalisms in order to investigate  phase transition of a charged AdS black hole with a global monopole. Here, we calculate   the curvature scalar of Weinhold and  
Ruppeiner metrics and the respective plots suggest that the curvature scalar of Weinhold and Ruppeiner metrics has one singular point, which   coincides only with zero point of the heat capacity (physical limitation point). We also apply the Quevedo and HPEM methods to investigate the thermodynamic properties of a charged AdS black hole with a global monopole.
Here we find that  the curvature scalar of Quevedo  metric (case-I) has three singular points, which coincide with zero point (physical limitation point) and divergence points (transition critical points) of heat capacity, respectively. However, the curvature scalar of Quevedo (case-II) metric has three singular points, in which two of them only coincide with divergences points (transition critical points) of heat capacity and no point coincides with zero point of heat capacity.
In case of HPEM metric, we find that the divergence points of the Ricci scalar  are coincident with zero point (physical limitation point) and divergence points (transition critical points) of heat capacity, respectively. So, the divergence points of the Ricci scalar of HPEM and Quevedo (case-I) metrics coincide with both types of phase transitions of the heat capacity.

The paper is presented systematically in following manner. In section \ref{section3}, we
recapitulate the basic setup of charged AdS black hole with global monopole.
In section \ref{sub},  we discuss the thermodynamics of charged AdS black hole with global monopole. Here, we do graphical analysis of themodynamical variables, in particular,
temperature, heat capacity and $P-v$ diagrams. The thermodynamic geometry of the system
with various formalisms are presented in section \ref{sub1}. Final remarks are
given in section \ref{section4}.
 
\section{The metric of a charged AdS black hole with global monopole}\label{section3}
In this section, we briefly review the metric components in the context of a charged AdS black hole with global monopole.  Let us first consider the Lagrangian density that characterizes the  charged AdS black hole  with a global monopole which is given by
\begin{equation}
\mathcal{L} = R-2\Lambda +\frac{1}{2}{\partial _\mu }{\phi ^a}{\partial ^\mu }{\phi ^ * }^a - \frac{\gamma }{4}{({\phi ^a}{\phi ^ * }^a - {\eta_{0} ^2})^2},
\end{equation}
where $R$ is the Ricci scalar, $\Lambda$ is the cosmological constant, $ \gamma $ is a constant, $\phi^{a} (a=1,2,3)$ is a triplet of scalar field, and $ \eta_{0} $ is the energy scale of symmetry breaking. 
The field configuration of  scalar field triplet is given by
\begin{eqnarray}
\phi ^a=\eta_0 h(\tilde r) \frac{\tilde x^a}{\tilde{r}},
\end{eqnarray}
where $\tilde x^a\tilde x^a = \tilde r^2$.
In general, a static spherically symmetric metric of a  charged AdS black hole with global monopole, can be given by \cite{Barriola:1989hx} 
 \begin{equation}\label{dtildas}
 d{{s}^2} =  - \tilde{f}(\tilde{r})d{\tilde{t}^2} + \tilde{f}{(\tilde r)^{ - 1}}d{ \tilde r^2} + { \tilde  r^2}(d{\theta ^2} + \sin^{2} \theta d{\varphi ^2}).
 \end{equation}
 The field equation  for  scalar field $\phi^a$ in the metric (\ref{dtildas})  is obtained by
 \begin{eqnarray}
 \tilde{f}h''+2\tilde{f}\frac{h'}{\tilde{r}}+\tilde{f}'h'-2\frac{h}{\tilde{r}^2}-\gamma \eta_0^2 h(h^2-1)=0.
 \end{eqnarray}
In certain approximation, the solution for  a charged AdS black hole with global monopole
is given by \cite{deng} 
\begin{equation}\label{dtildaf}
\tilde f(\tilde r)  = 1 - 8\pi {\eta_{0} ^2} - \frac{{2\tilde m}}{\tilde r} + \frac{{{\tilde q^2}}}{{{\tilde r^2}}} + \frac{{{\tilde r^2}}}{{{l^2}}}.
\end{equation}
in which $l$, $\tilde  m $ and $\tilde q$  are AdS radius related to the cosmology constant as $l^{2} =-\dfrac{3}{\Lambda} $, the mass and electric charge of the black hole, respectively.
Due to this metric,  thermodynamical quantities of the black holes exhibit
an interesting dependence on the internal global monopole, and they
perfectly satisfy both the first law of thermodynamics and Smarr relation \cite{deng}.
 In following, using coordinate transformations \cite{Barriola:1989hx,Kumara:2019xgt,AhmedRizwan:2019yxk,Gunasekaran:2012dq}
 \begin{equation}
\tilde{t} = {(1 - 8\pi {\eta_{0} ^2})^{ - \frac{1}{2}}}t    ,      \;\; \;\;    \;\;\;\; \;\;       \tilde{r} = {(1 - 8\pi {\eta_{0} ^2})^{\frac{1}{2}}}{r},
\end{equation}
and also, with the help of new parameters
 \begin{equation}
m = {(1 - 8\pi {\eta_{0} ^2})^{ - \frac{3}{2}}} \tilde{m}   ,      \; \;\;   \; \;\;       q = {(1 - 8\pi {\eta_{0} ^2})^{-1}} \tilde{q} ,     \;\;\;\; \;\; \eta ^2=8\pi {\eta_{0} ^2},
\end{equation}
we have the line element (\ref{dtildas}) 
 \begin{equation}
d{s^2} =  - f(r)d{t^2} + f{(r)^{ - 1}}d{r^2} + (1-{\eta ^2}) {r^2}(d{\theta ^2} + \sin^{2} \theta d{\varphi ^2}),
\end{equation}
where 
\begin{equation}
f(r) = 1 - \frac{{2m}}{r} + \frac{{{q^2}}}{{{r^2}}} + \frac{{{r^2}}}{{{l^2}}}.
\end{equation}
Here, the spacetime described by this metric exhibits a solid angle deficit.
Also, the electric charge and the ADM mass  are given by \cite{Barriola:1989hx,Kumara:2019xgt,AhmedRizwan:2019yxk,Gunasekaran:2012dq}
\begin{equation}
Q = (1 - {\eta ^2}) q  \;\;\; \mbox{and}\; \;   M = (1 - {\eta ^2}) m.
\end{equation}
Here, we conclude that the thermodynamical quantities of the black holes depend on the internal global monopole.
\section{Thermodynamics}\label{sub}
\begin{figure}[h]
	\centering
	\subfigure[$\eta=0.5, l=5.263, l=0.5 $]{
		\includegraphics[width=0.38\textwidth]{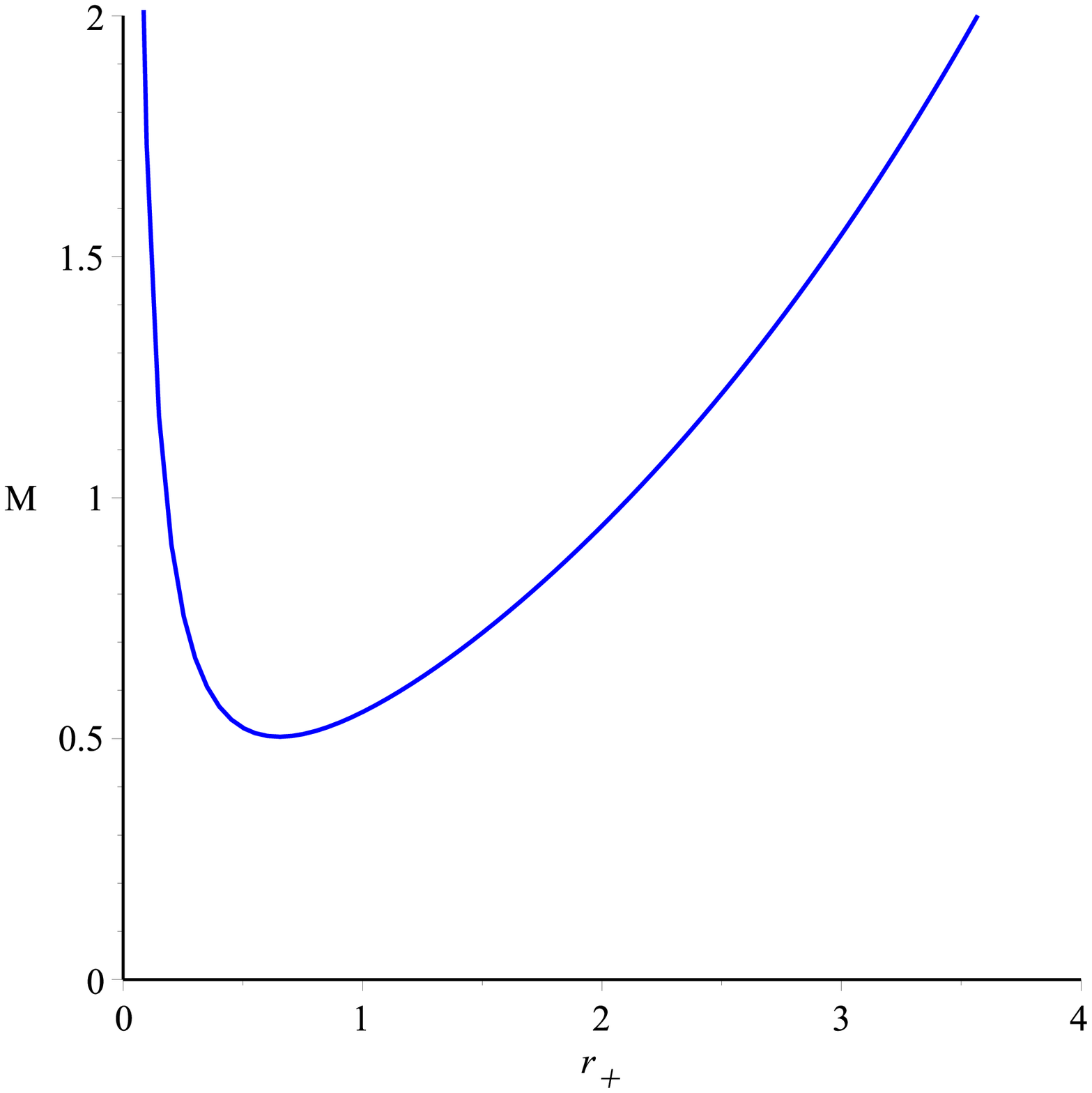}
	}
\subfigure[$\eta=0.5,  l=5.263 $]{
	\includegraphics[width=0.38\textwidth]{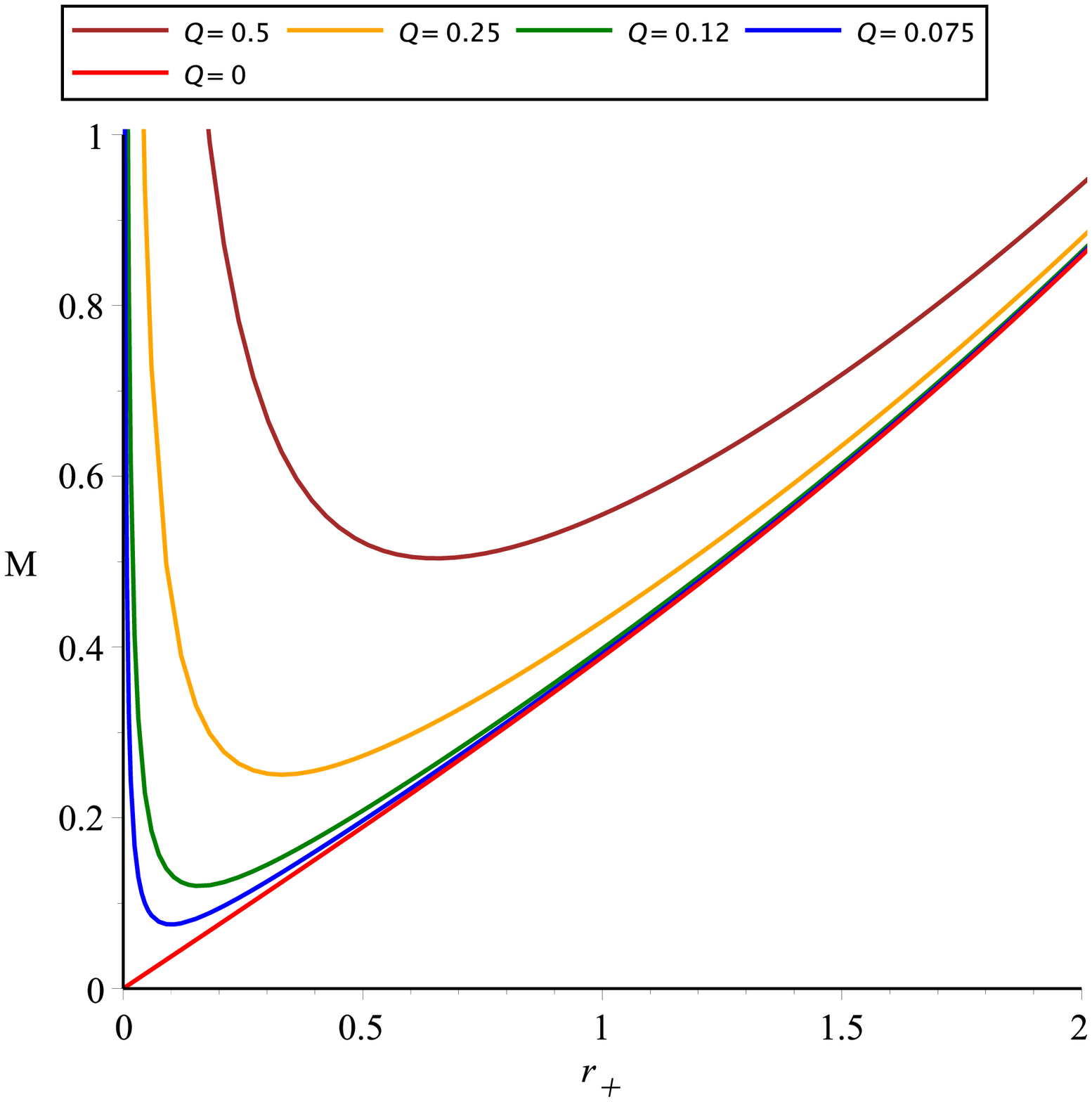}
}
	\subfigure[$\eta=0.5, Q=0.5 $]{
		\includegraphics[width=0.38\textwidth]{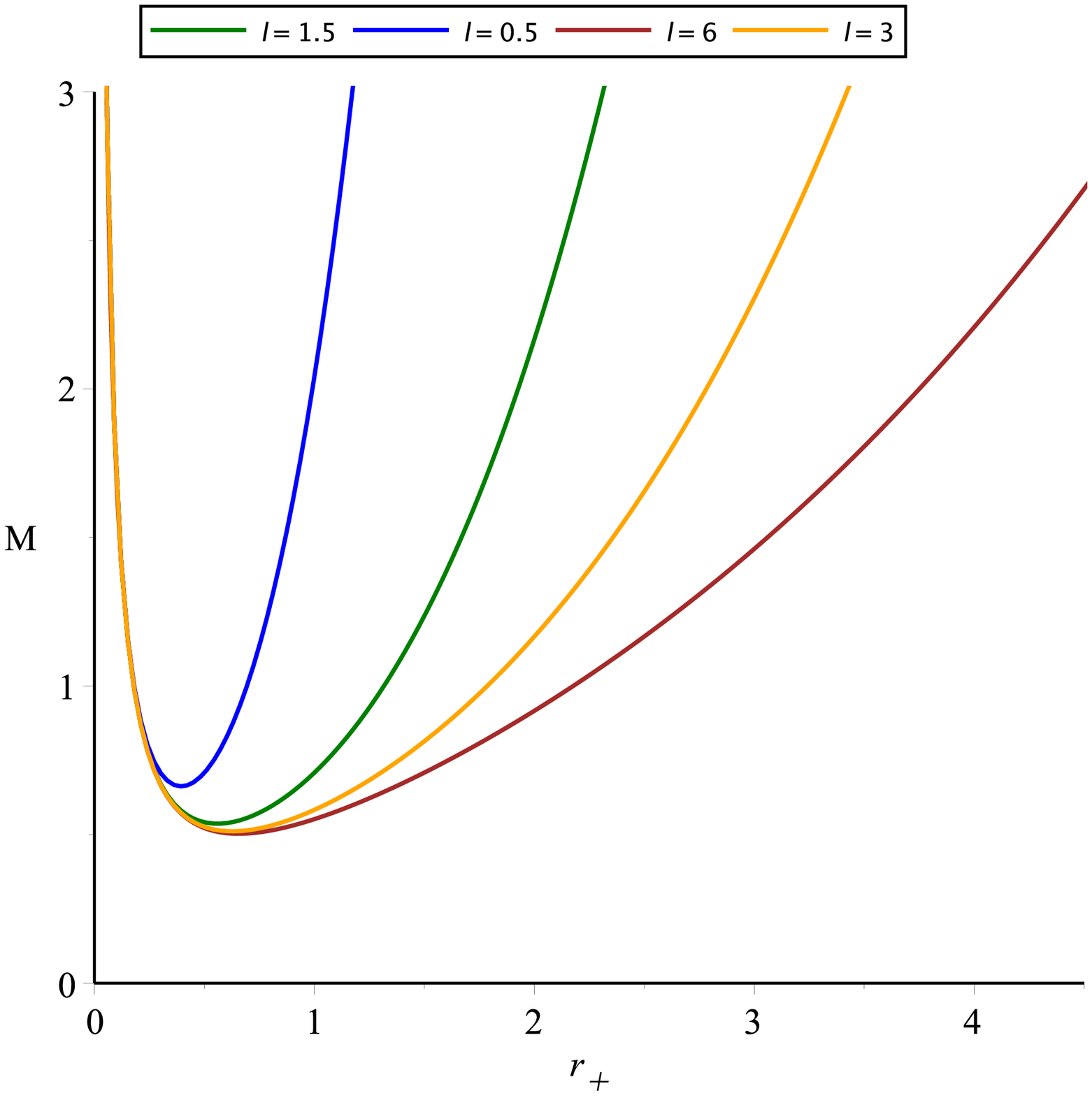}
	}
	\subfigure[$Q=0.5, l=5.263 $]{
		\includegraphics[width=0.38\textwidth]{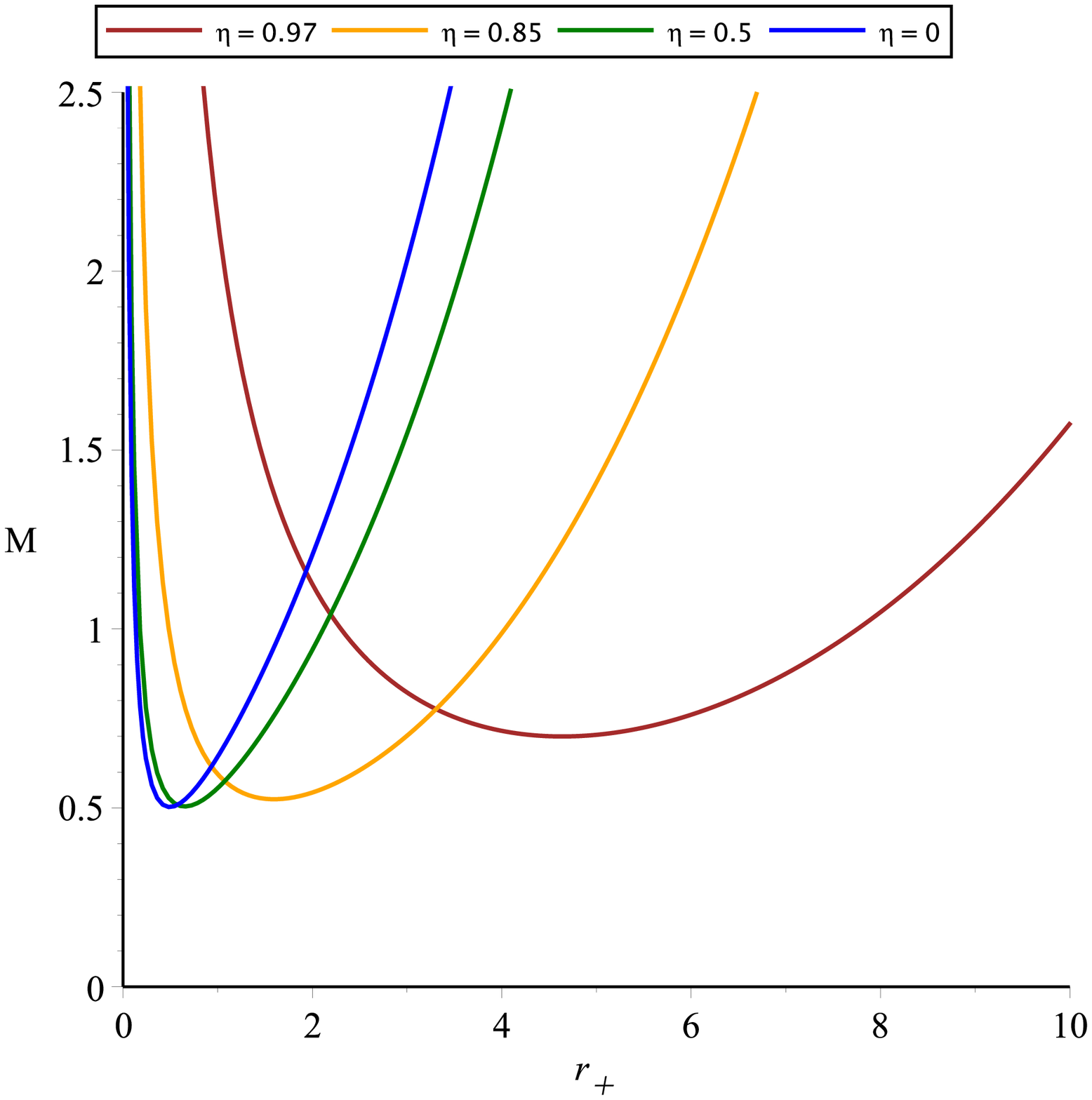}
	}
	\caption{Variations of mass in terms of horizon radius $ r_{+}$.}
	\label{pic:M}
\end{figure}

\begin{figure}[h]
	\centering
	\subfigure[$\eta=0.5, l=5.263, Q=0.5 $]{
		\includegraphics[width=0.38\textwidth]{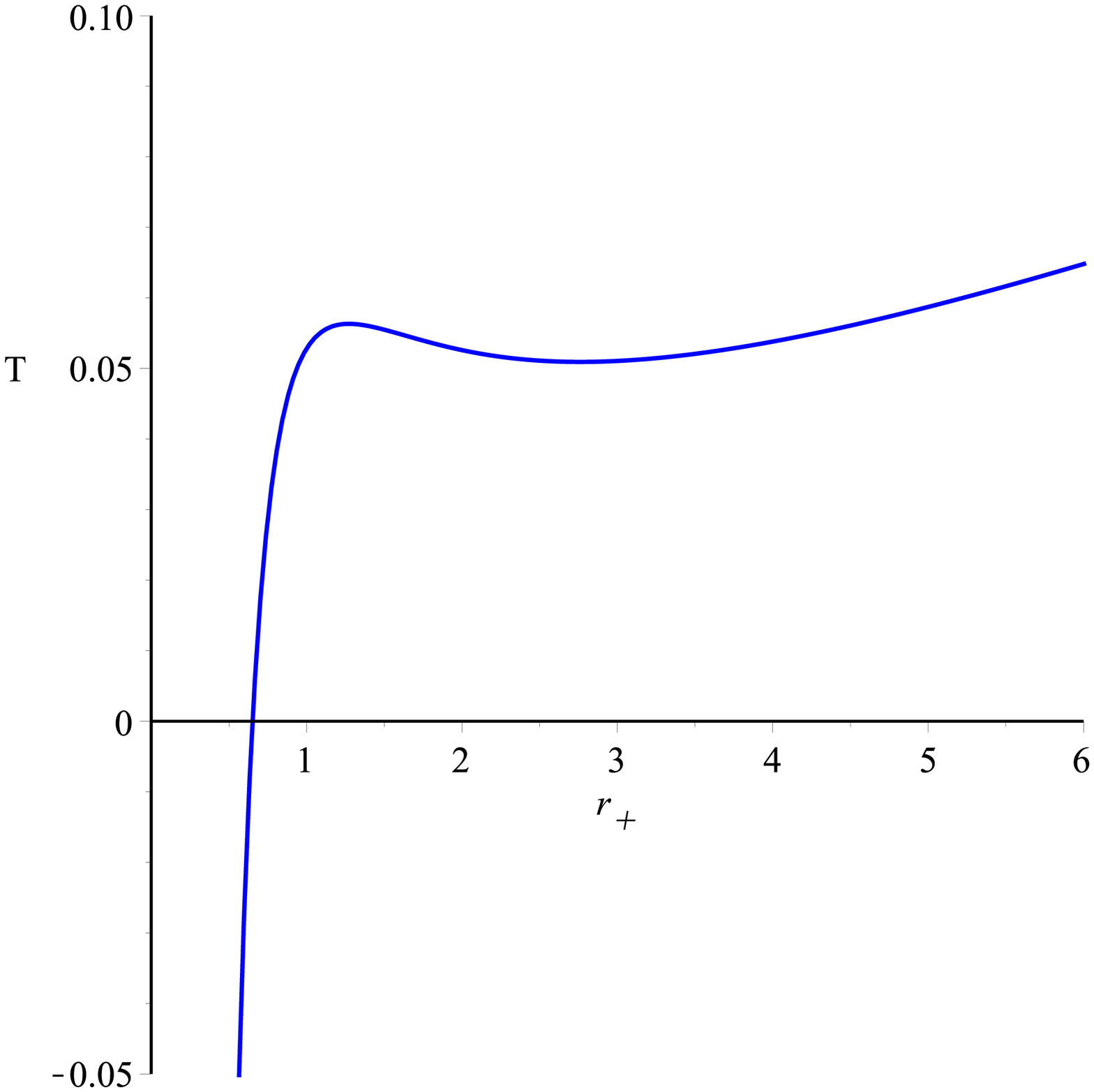}
	}
\subfigure[$\eta=0.5, Q=0.075 $]{
	\includegraphics[width=0.38\textwidth]{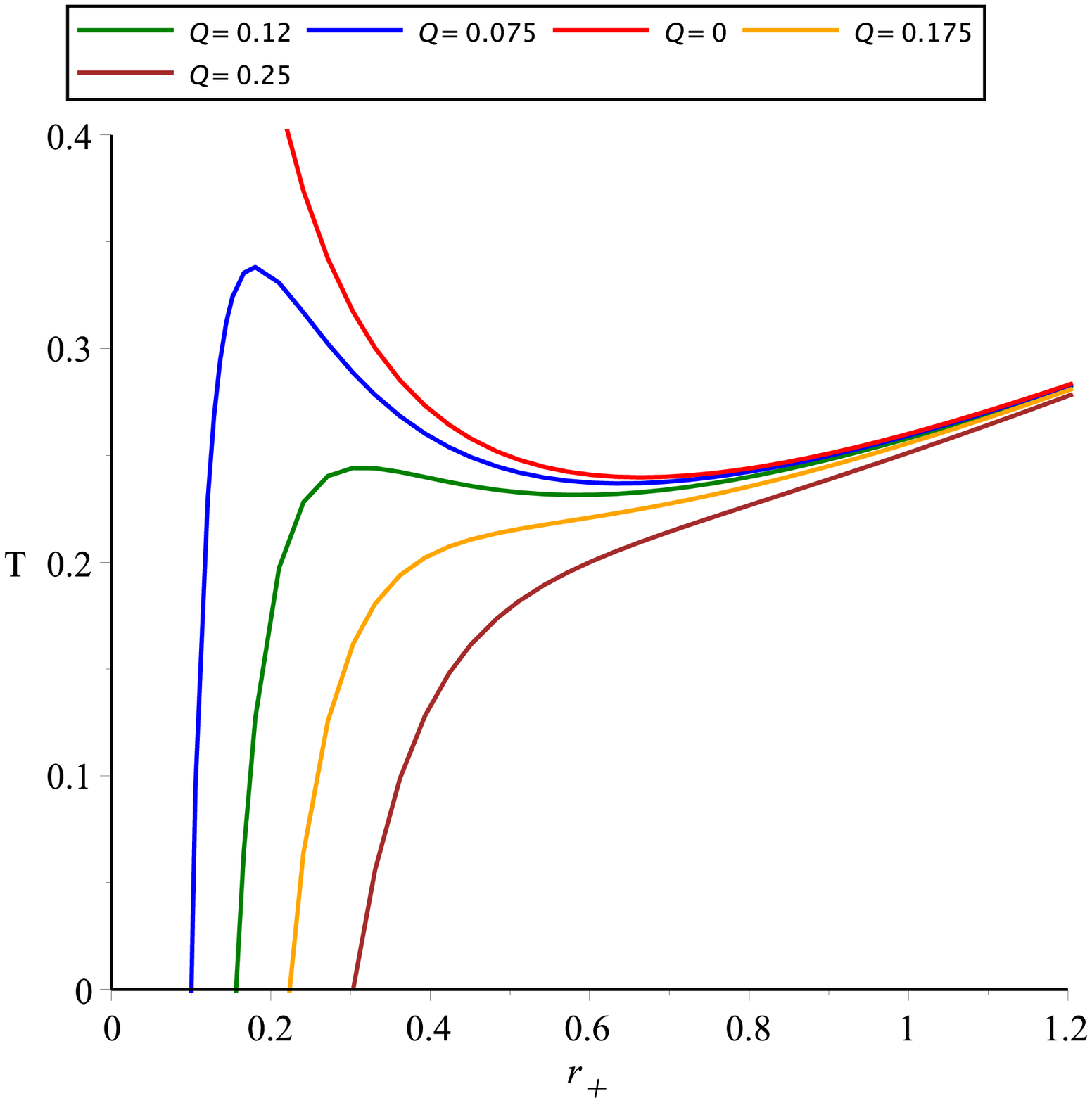}
}
	\subfigure[$\eta=0.5, Q=0.075 $]{
		\includegraphics[width=0.38\textwidth]{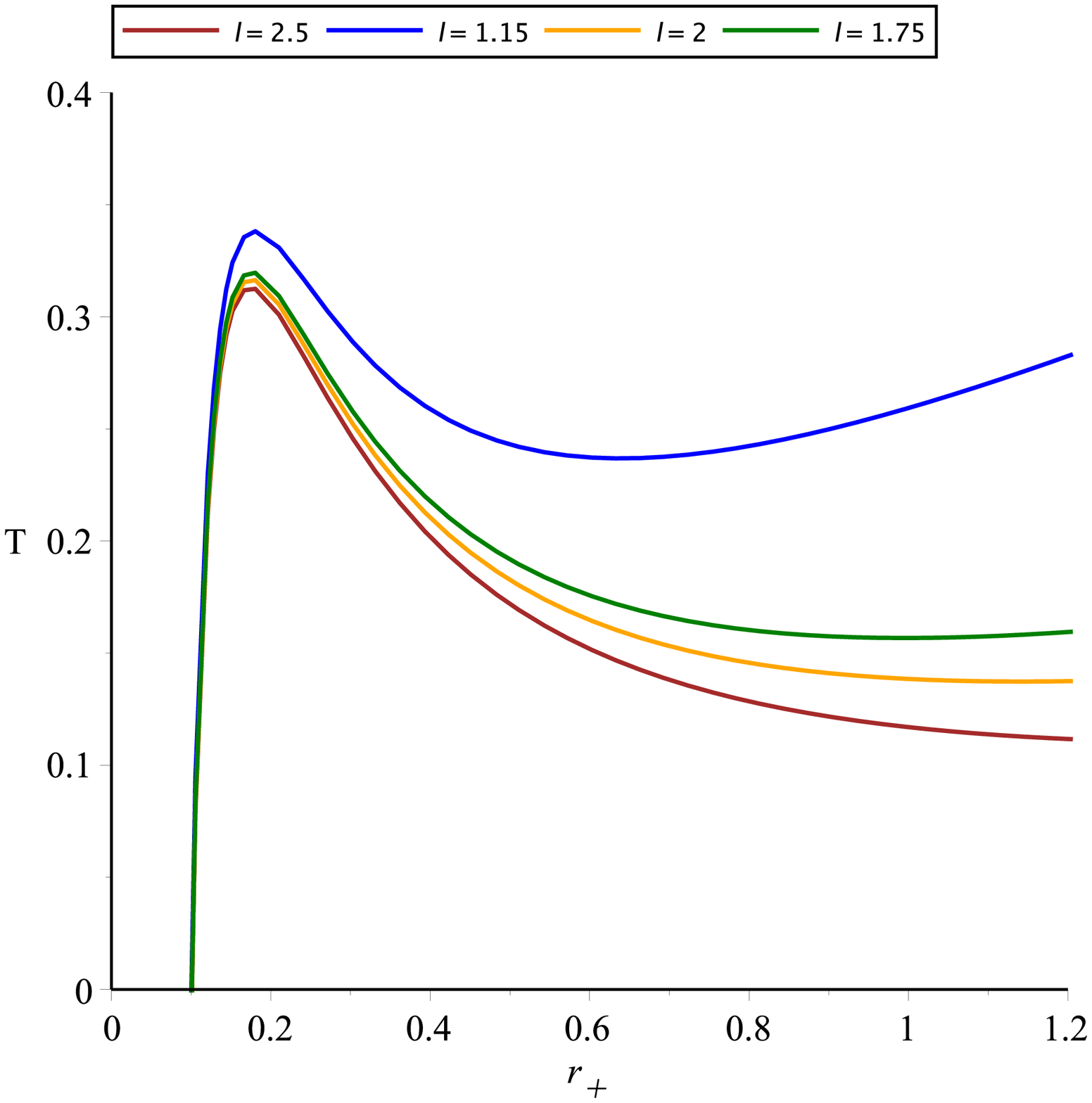}
	}
	\subfigure[$Q=0.075, l=1.15 $]{
		\includegraphics[width=0.38\textwidth]{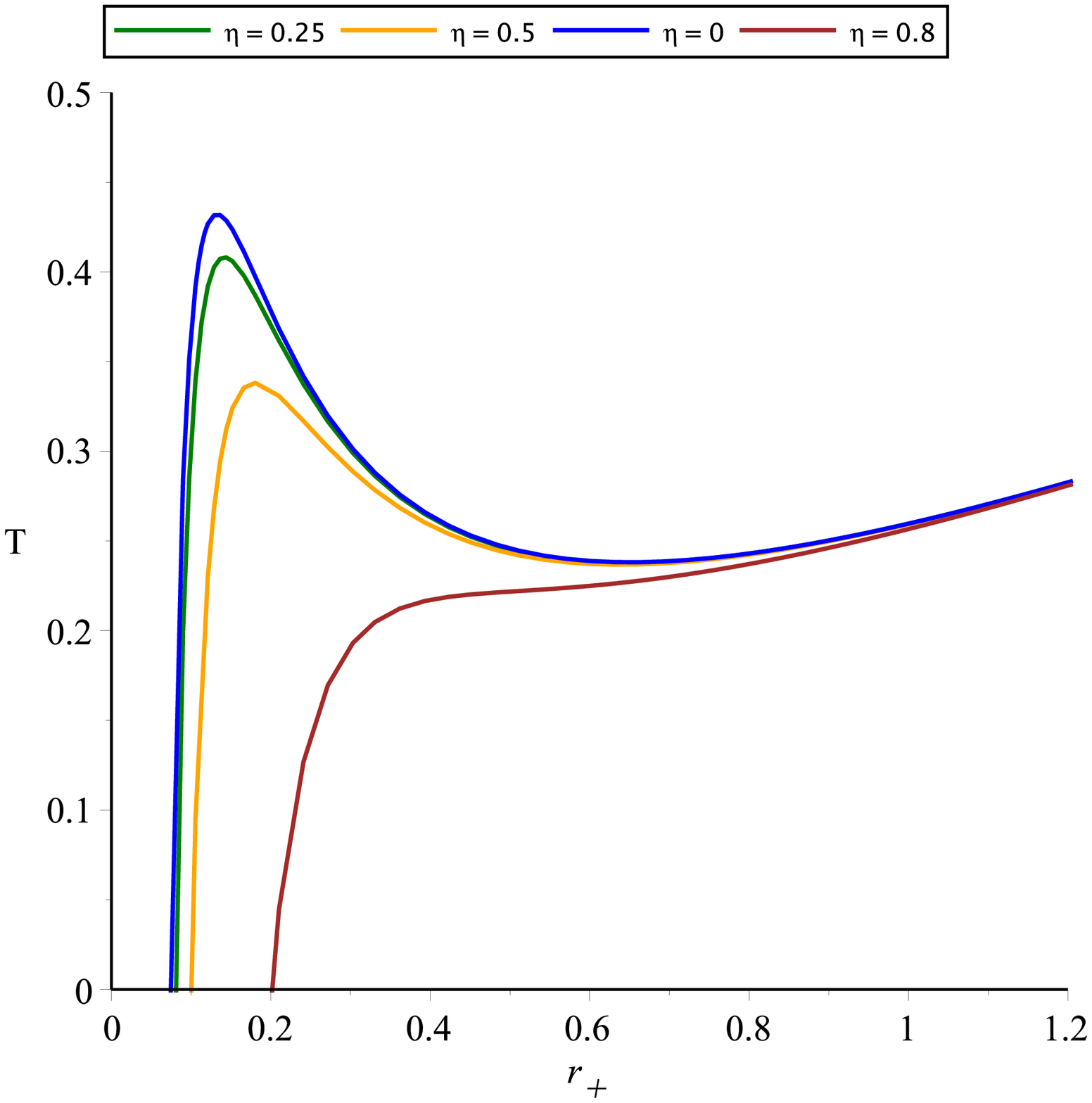}
	}
	\caption{Variations of temperature in terms of horizon radius $ r_{+}$.}
	\label{pic:T}
\end{figure}

\begin{figure}[h]
	\centering
		\subfigure[$Q=0.5,\eta=0.5, l=5.263 $]{
		\includegraphics[width=0.3\textwidth]{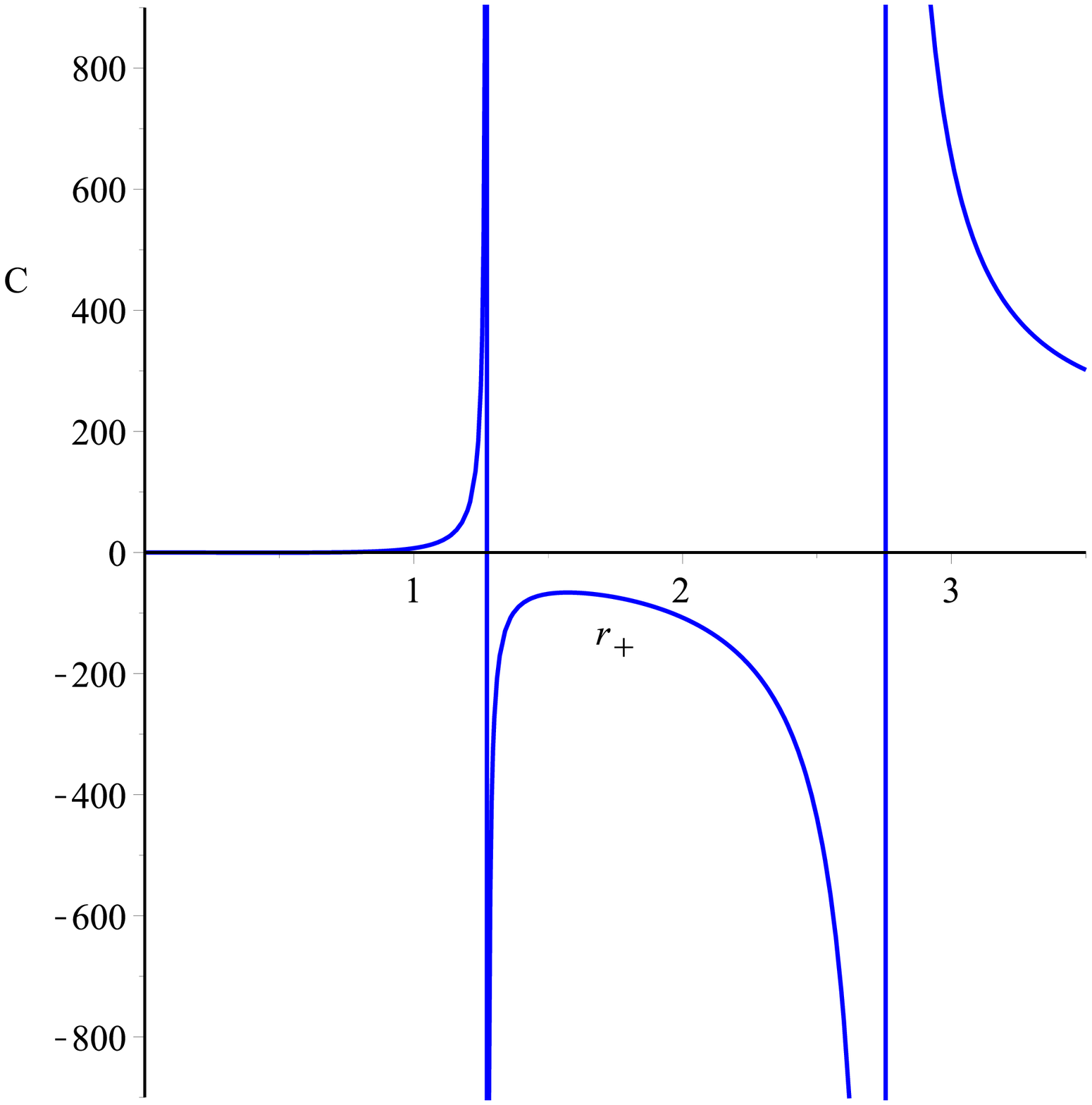}
	}
\subfigure[closeup of figure (a)]{
	\includegraphics[width=0.3\textwidth]{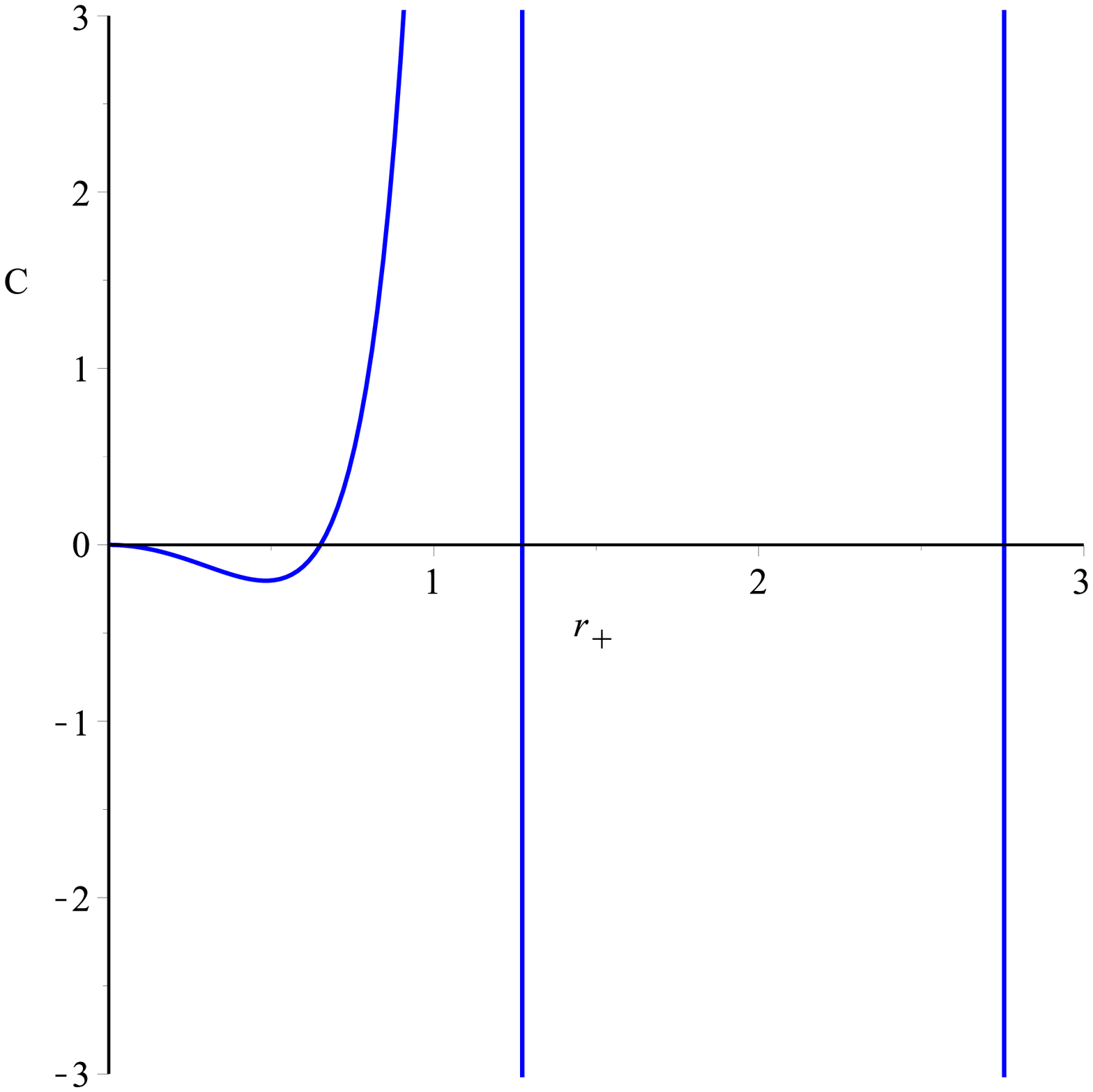}
}
	\subfigure[$\eta=0.5, l=1.15 $]{
		\includegraphics[width=0.3\textwidth]{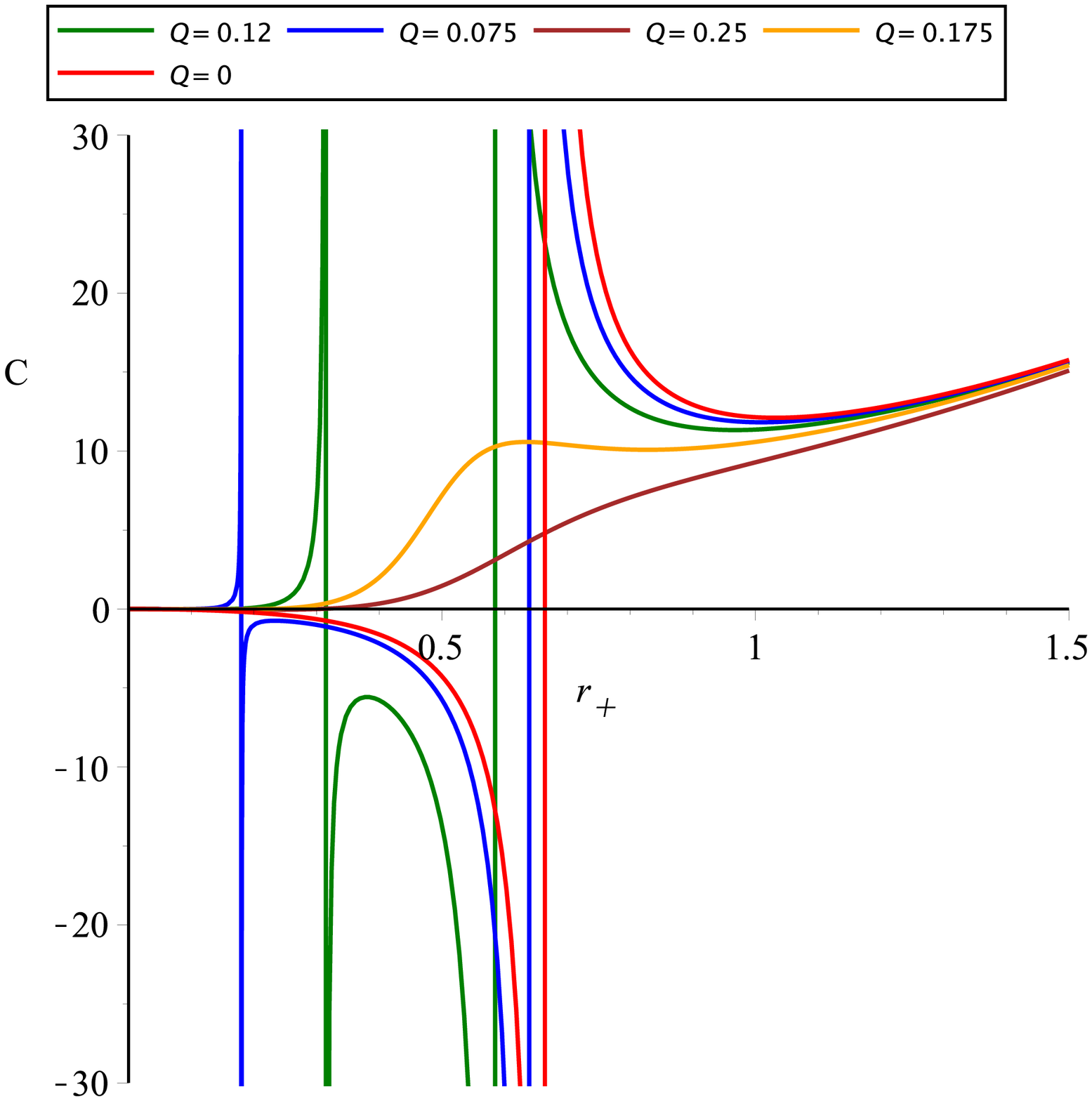}
	}
\subfigure[closeup of figure (b) ]{
	\includegraphics[width=0.3\textwidth]{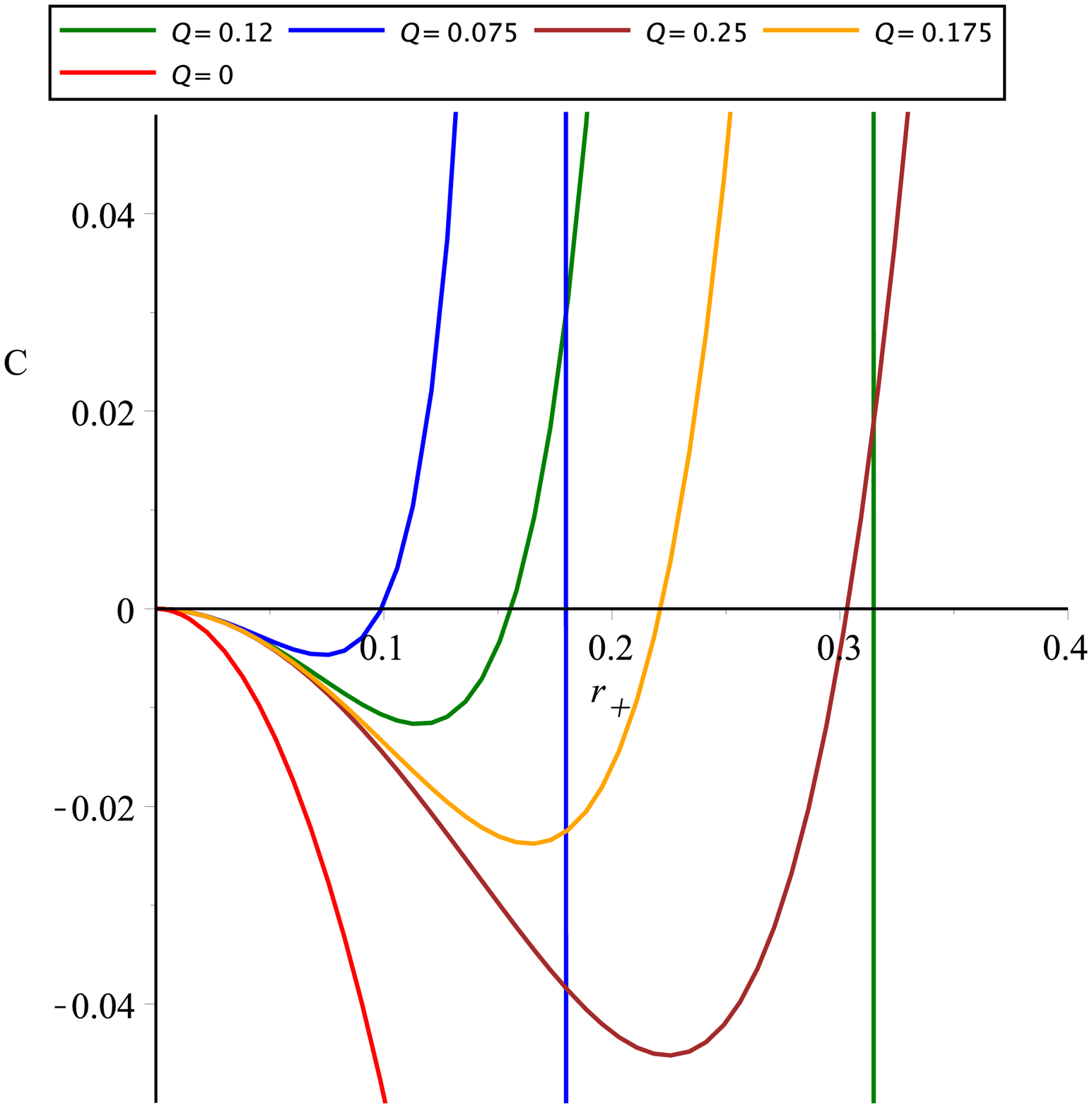}
}
	\subfigure[$\eta=0.5, Q=0.075 $]{
		\includegraphics[width=0.3\textwidth]{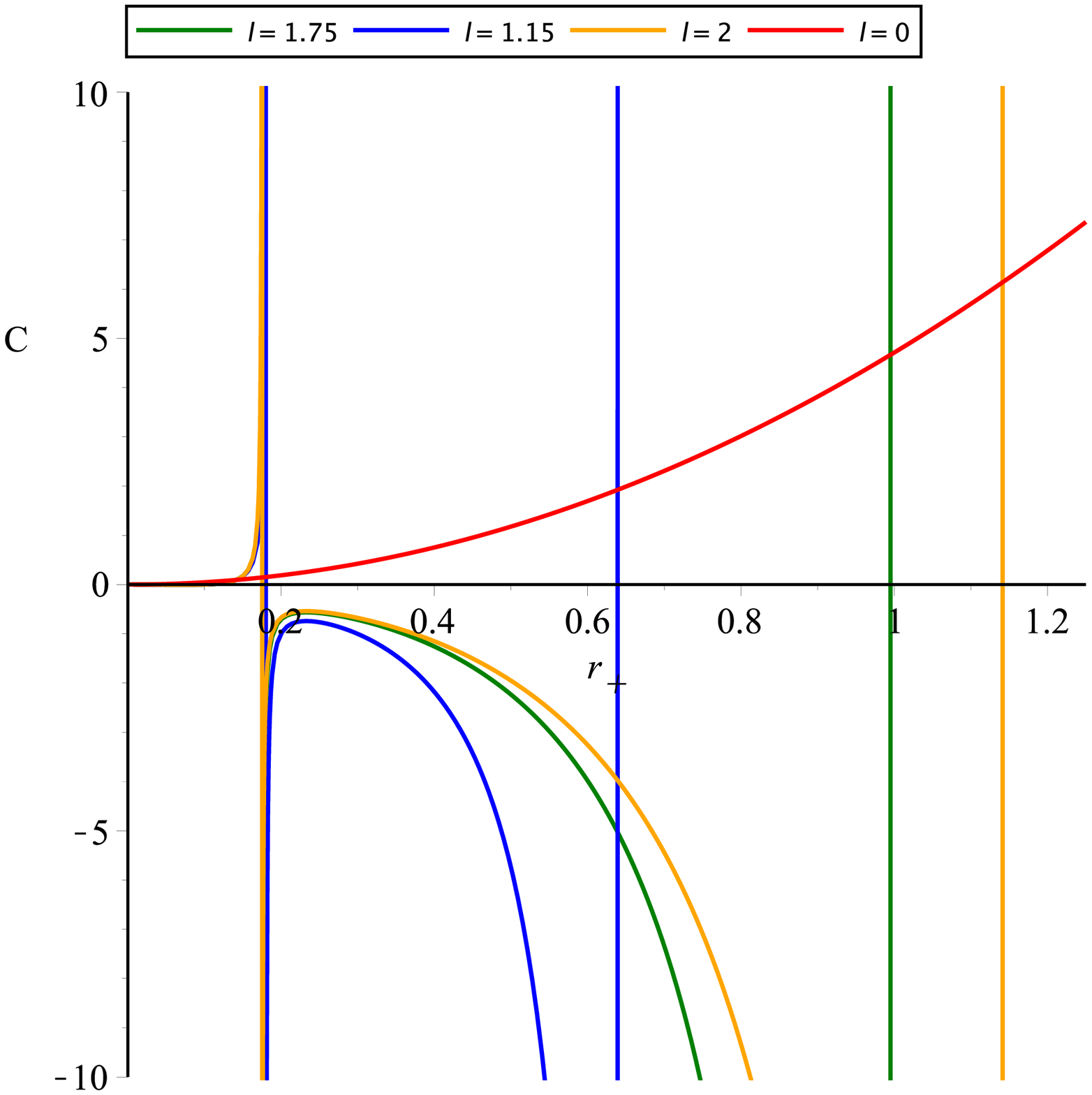}
	}
\subfigure[closeup of figure (e) ]{
	\includegraphics[width=0.3\textwidth]{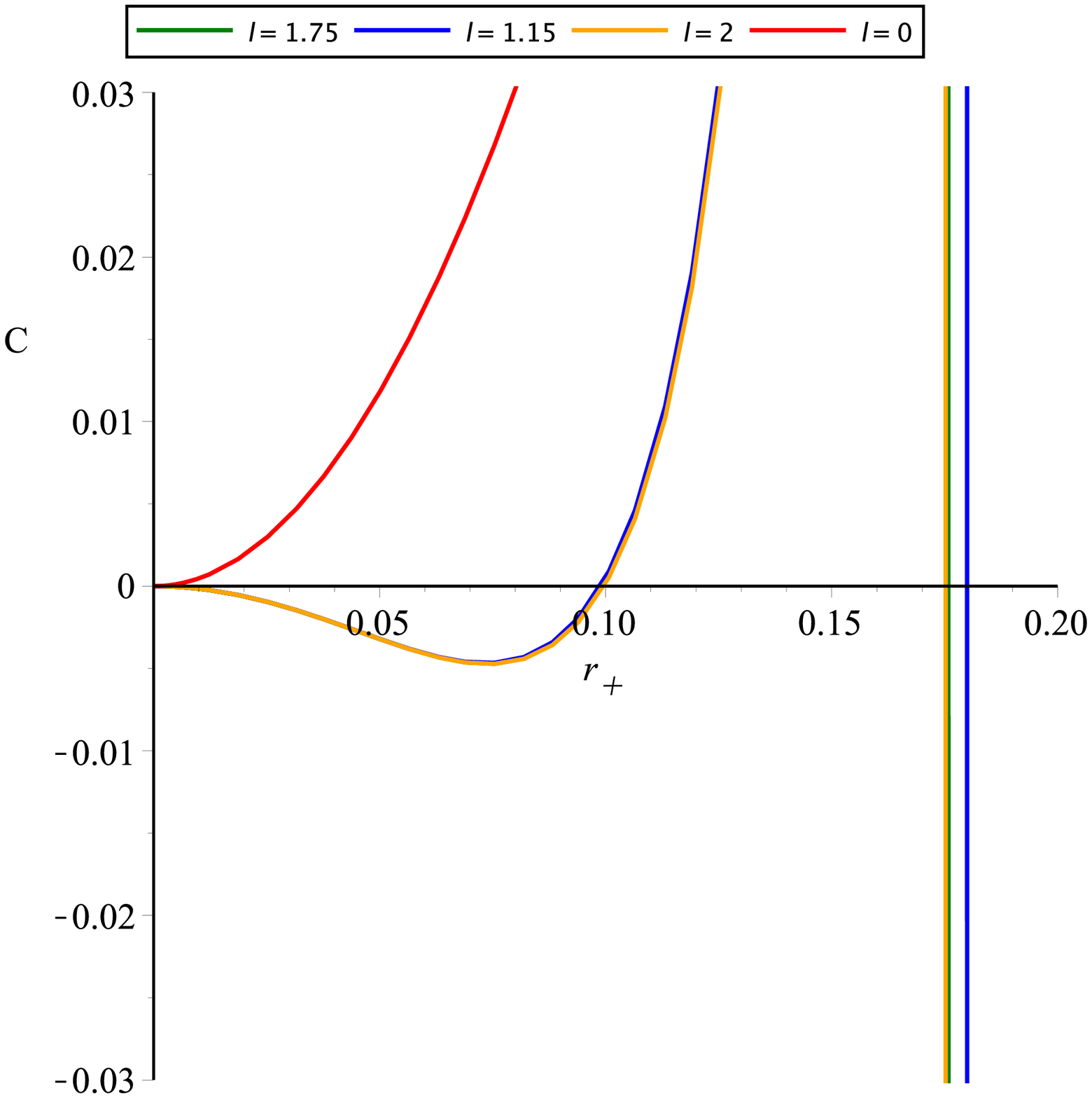}
}
	\subfigure[$Q=0.075, l=1.15 $]{
		\includegraphics[width=0.3\textwidth]{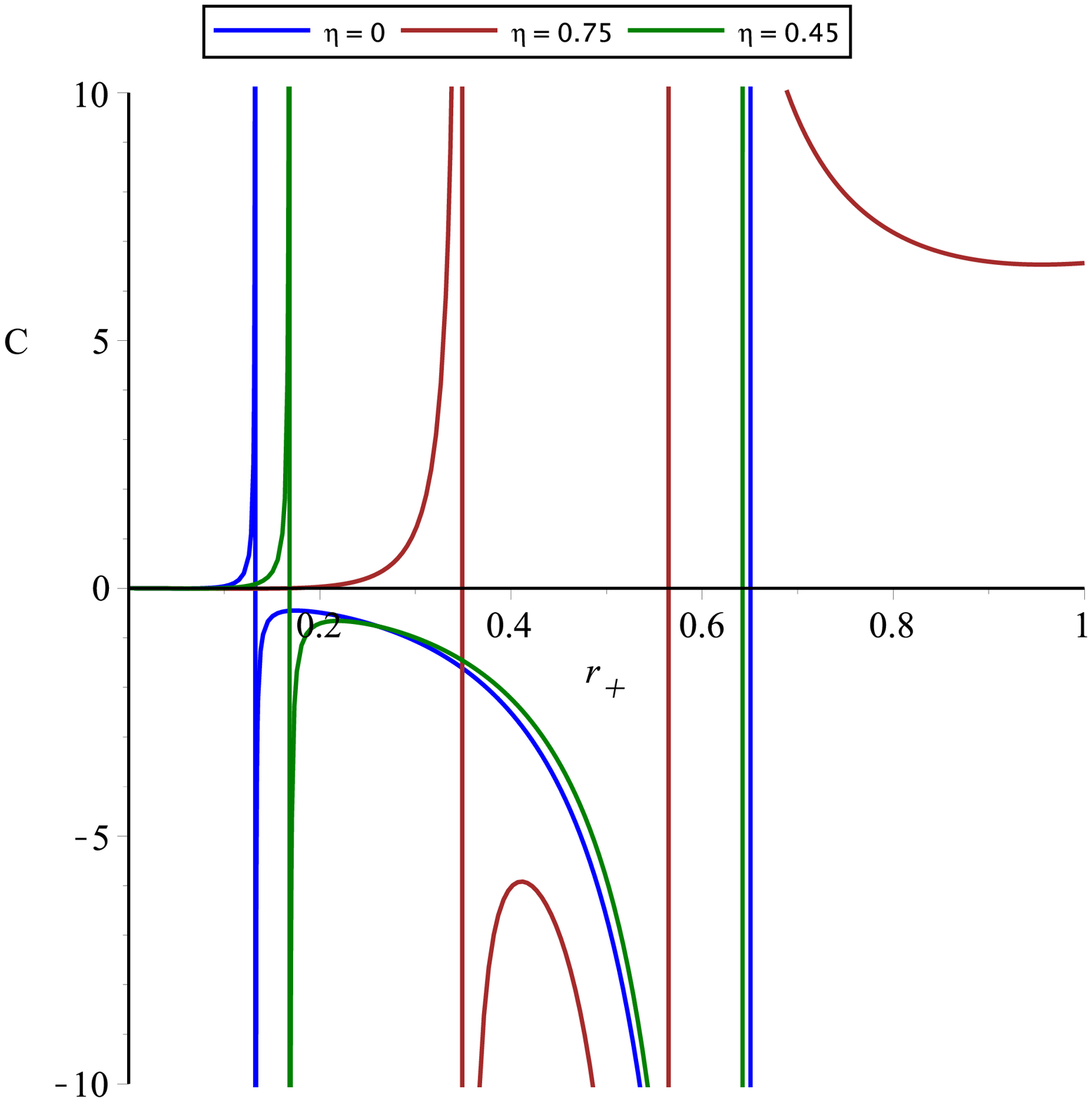}
	}
\subfigure[closeup of figure (g) ]{
	\includegraphics[width=0.3\textwidth]{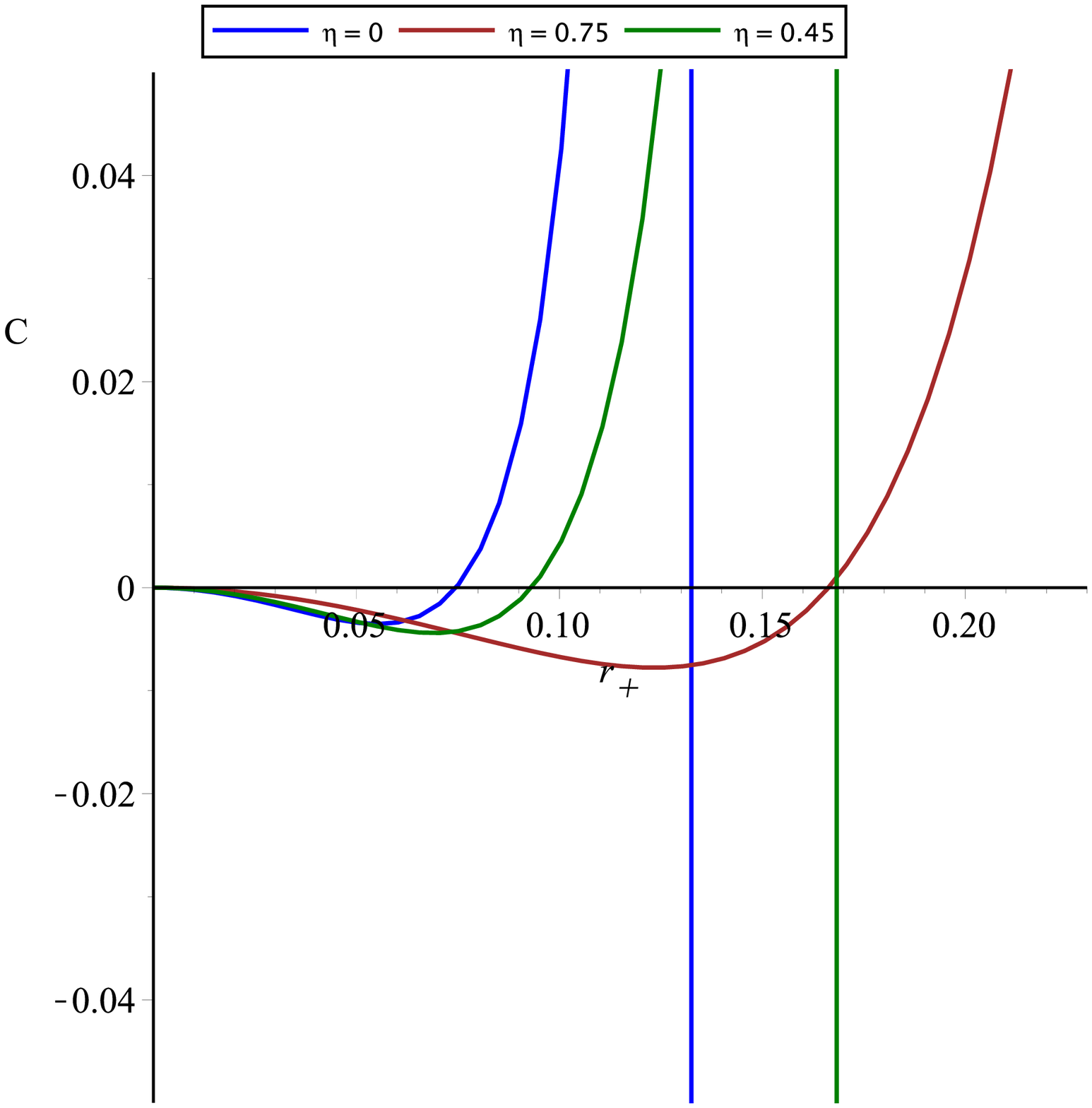}
}
	\caption{Variations of the heat capacity in terms of horizon radius $ r_{+}$.}
	\label{pic:C}
\end{figure}

\begin{figure}[h]
	\centering
	\subfigure[$\eta=0 $]{
		\includegraphics[width=0.38\textwidth]{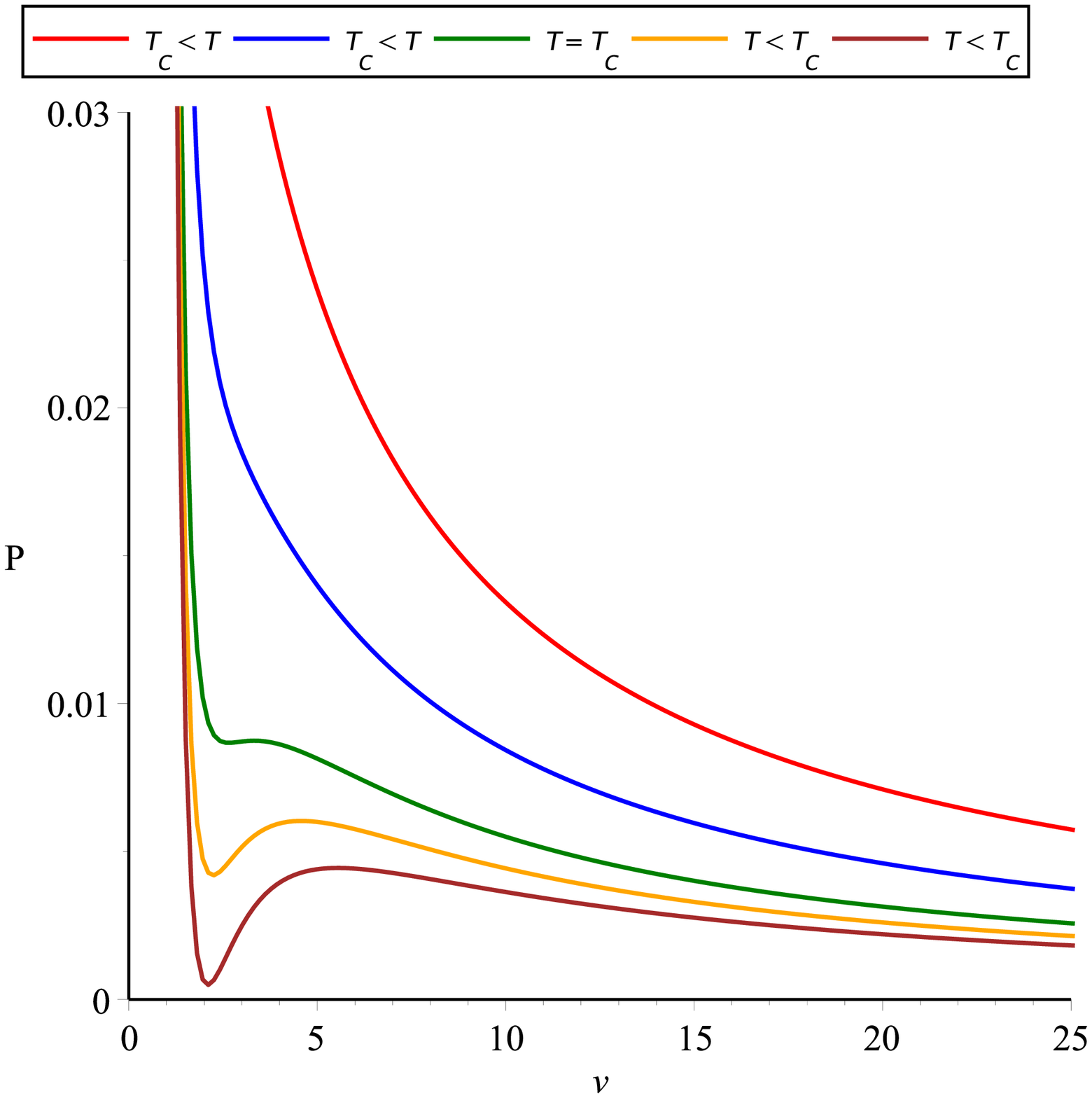}
	}
	\subfigure[$\eta=0.5 $]{
		\includegraphics[width=0.38\textwidth]{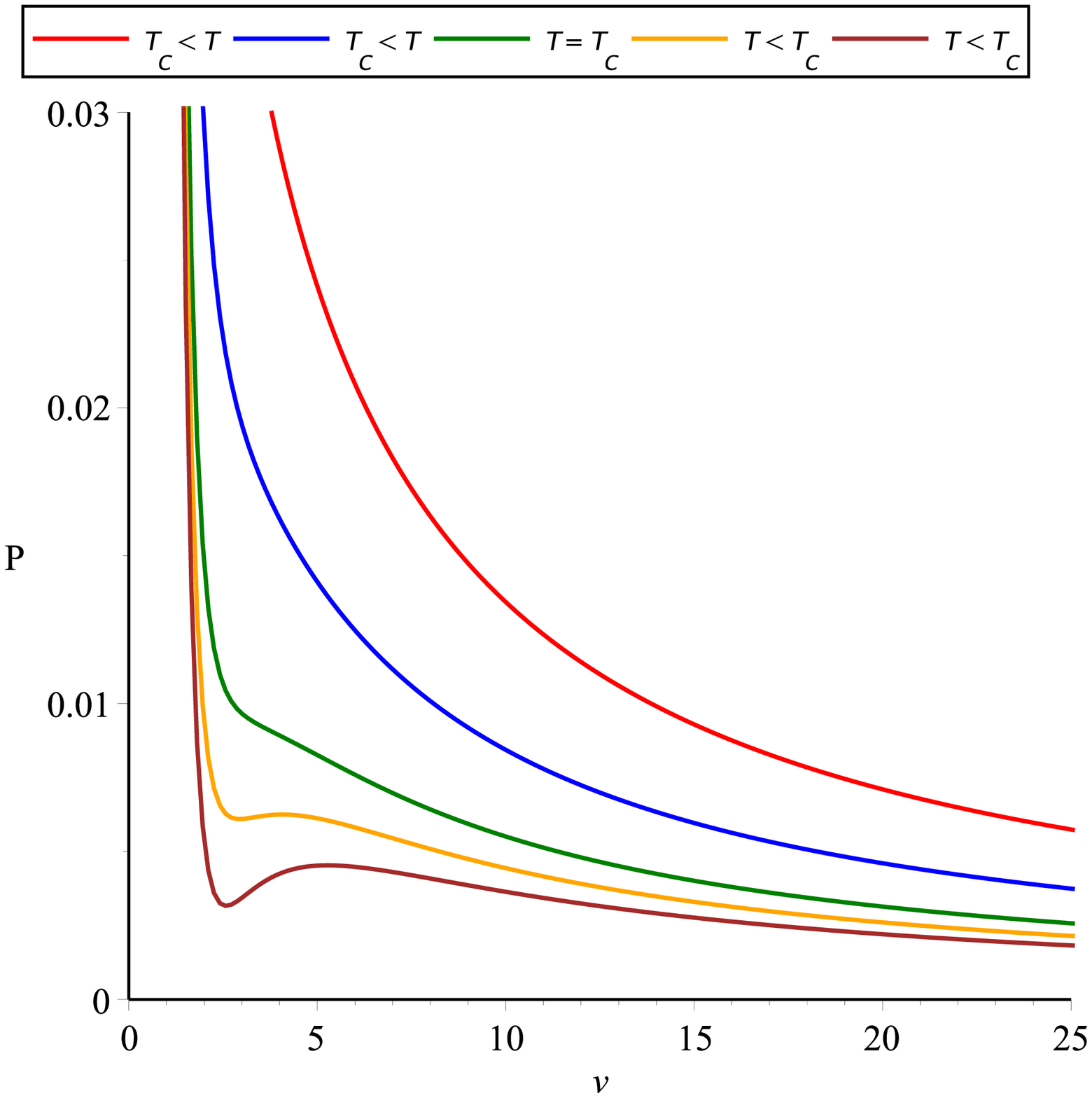}
	}
\subfigure[$\eta=0.75$]{
	\includegraphics[width=0.38\textwidth]{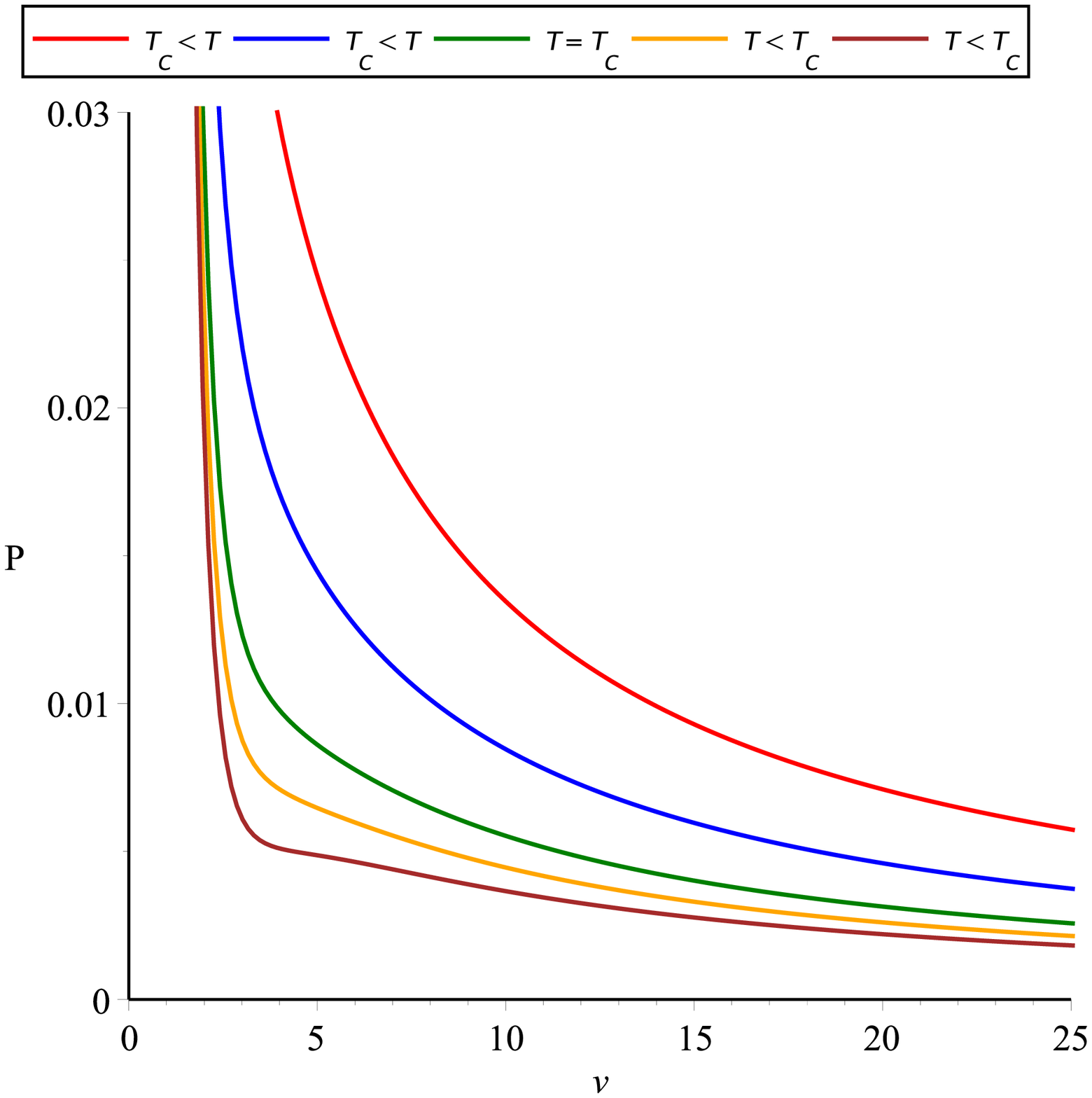}
}
	\subfigure[$\eta=0.97 $]{
		\includegraphics[width=0.38\textwidth]{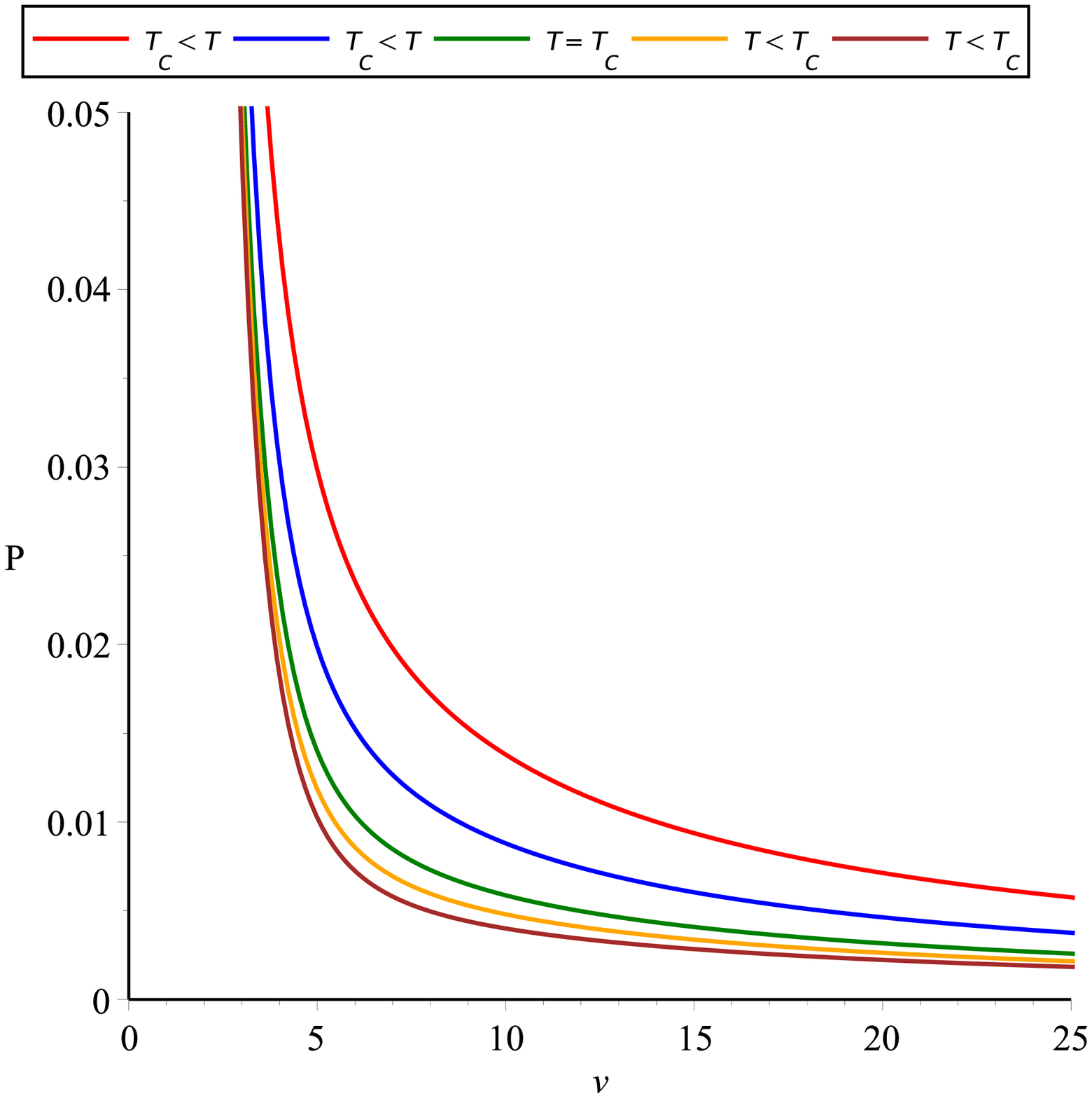}
	}
	\caption{$P$ - $v$ diagram of a charged AdS black hole with a global monopole in extended phase space for  $ Q = 0.85$, $l=1.5$ and different values of $\eta$.}
	\label{pic:P}
\end{figure}

To determine the mass parameter, we use the condition  $f(r_{+})=0$, at the event horizon $ r = r_{+}$, and express it in terms of entropy $S$, using the relation between entropy $S$  and event horizon radius
$r_{+}$  $ (S=\pi  (1 - {\eta ^2}) r^{2}_{+})$. This gives 
\begin{equation}\label{RQmass}
	M(S,Q,l)=-\dfrac{1}{2}\dfrac{\pi ^2 Q^2 l^2-\pi S \eta ^2 l^2+\pi S l^2+S^2}{{l^2}{\pi ^{3/2}}\left( {{\eta ^2} - 1} \right)\left( {\sqrt { - \frac{S}{{{\eta ^2} - 1}}} } \right)}  .
\end{equation}
In the extended phase space, the first law of thermodynamics for a charged AdS black hole with global monopole can be written as \cite{Kumara:2019xgt,AhmedRizwan:2019yxk,Gunasekaran:2012dq}
\begin{equation}\label{RQlow}
	{\it dM}={\it TdS}+\varphi {\it dQ}+PdV,
\end{equation}
and then, the Hawking temperature  $ T=\frac{\partial M}{\partial S}$, heat capacity $C=T\frac{\partial S}{\partial T}$, and electric potential $\varphi=\frac{\partial M}{\partial Q}$ can be obtained as 
\begin{eqnarray}\label{T}
	T&=&\dfrac{1}{4}{\frac {{\pi }^{2}{Q}^{2}{l}^{2}+\pi \,S{\eta}^{2}{l}^{2}-\pi \,S{l}^{2}-3\,{S}^{2}}{S{l}^{2} \left( {\eta}^{2}-1 \right)\pi^{\frac{3}{2}} } \left(  \sqrt{-{\frac {S}{{\eta}^{2}-1}}} \right) ^{-1}},\\
	C&=& 2  \frac { S\left( {\pi }^{2}{Q}^{2}{l}^{2}+\pi \,S{\eta}^{2}{l}^{2}-\pi \,S{l}^{2}-3\,{S}^{2} \right)    }{3\,{\pi }^{2}{Q}^{2}{l}^{2}+\pi \,S{\eta}^{2}{l}^{2}-\pi \,S{l}^{2}+3\,{S}^{2}} ,\\
 \varphi &=&\frac {\sqrt{\pi }Q}{{\eta}^{2}-1} \left(\sqrt{-{\frac {S}{{\eta}^{2}-1}}} \right) ^{-1} .
\end{eqnarray}
Moreover, in the extended phase space, the cosmological constant is denoted as a thermodynamic pressure \cite{Kumara:2019xgt,AhmedRizwan:2019yxk,Gunasekaran:2012dq}, 
\begin{equation}\label{P1}
	P=-{\frac {\Lambda}{8 \pi }}={\frac {3}{8\pi \,{l}^{2}}}  ,
\end{equation}
and its conjugate quantity associates to the thermodynamic volume as \cite{Kumara:2019xgt,AhmedRizwan:2019yxk,Gunasekaran:2012dq}
\begin{eqnarray}
V&=&\frac{{\partial M}}{{\partial P}}=-\dfrac{4}{3}{\frac {{S}^{2}}{ \sqrt{\pi } \left( {\eta}^{2}-1 \right) } \left(  \sqrt{-{\frac {S}{{\eta}^{2}-1}}} \right) ^{-1}}\nonumber\\
&=&\dfrac{4}{3}\pi \,{r_{+}}^{3} \left( {\eta}^{2}-1 \right)  .
\end{eqnarray}
Also, the equation of state can be yields with combining equations (\ref{T}) and (\ref{P1}) as \cite{Kumara:2019xgt,AhmedRizwan:2019yxk,Gunasekaran:2012dq}
\begin{eqnarray}
P&=&{\frac {T}{2 r_{+}}}-{\frac {1}{8 \pi \,{r_{+}}^{2}}}+{\frac {{Q}^{2}}{8 \pi \, \left(1-{\eta}^{2}\right) {r_{+}}^{4}}}\nonumber\\
&=&{\frac {T}{v}}-{\frac {1}{2\pi \,{v}^{2}}}+{\frac {{Q}^{2}}{\pi \, \left(1-{\eta}^{2}\right) {v}^{4}}} .
\end{eqnarray}
Here, the relation between the horizon radius and the specific volume as $ v=2 l^{2} r_{+} $
is utilized \cite{Kumara:2019xgt,AhmedRizwan:2019yxk,Gunasekaran:2012dq}.
The critical parameters can be obtained by utilizing
the vanishing derivatives at the critical point,
\begin{equation}
\left(\frac{{\partial P}}{{\partial v}}\right)_T = \left(\frac{{{\partial ^2}P}}{{\partial {v^2}}}\right)_T = 0  ,
\end{equation}
and therefore
\begin{equation}
T_c = \frac{{(1 - {\eta ^2})}}{{3\sqrt 6 \pi Q}} \quad ,\quad  {{\rm{P}}_c} = \frac{{{{(1 - {\eta ^2})}^2}}}{{96\pi {Q^2}}}\quad ,\quad v_c= \frac{{2\sqrt 6 Q}}{{(1 - {\eta ^2})}}  .
\end{equation}
To better study, these thermodynamic parameters ($M$, $T$, $C$ and $P$) are plotted versus horizon radius $r_+$, (see Figs.~\ref{pic:M},\ref{pic:T},\ref{pic:C} and \ref{pic:P}), and their behaviors are investigated.

The behavior of mass for different values of $Q$, $l$ and $\eta$, is shown in Fig.~\ref{pic:M}. It can be seen from Fig.~\ref{pic:M}(a), the mass of this black hole, has one minimum point  $r_+ = r_m $, in which the value of it, for $Q=0.5$, $l=5.263$ and $\eta=0.5$, is equal to 0.651. 
Moreover, from Fig. \ref{pic:M} (b)-(d), it is clear that by increasing the value of $Q$, $l$ and $\eta$, the  minimum point can be occur at different location.

Also, the behavior of temperature for different values of $Q$, $l$ and $\eta$, is demonstrated in Fig.~\ref{pic:T}. It can be observed from Fig.~\ref{pic:T}(a), the plot of the temperature is in the negative region at a particular range of $r_+(r_+< r_m)$, then it reaches to zero at $r_+=r_m$. Fig.~\ref{pic:T}(a)-(d), shows that for $r_+> r_m$, the temperature will be positive and increases to a maximum point.  After that it starts decreasing with higher $r_+$ and then again starts increases gradually. 
A comparative analysis of these plots 
reflects that as long as the
values of $Q$, $l$ and $\eta$ increases, the peak value of temperature becomes smaller.

In addition, the behavior of heat capacity for different values of $Q$, $l$ and $\eta$, is depicted in Fig.~\ref{pic:C}. Note that, the root of heat capacity ($C=T=0$) is showing a boundary line between non-physical ($ T < 0 $) and physical ($ T > 0 $) black holes and it call a physical limitation point \cite{EslamPanah:2018ums}. Sign of the heat capacity changes in this physical limitation point.   
As we can see from Fig. \ref{pic:C} (a), (b), for $Q=0.5$, $l=5.263$ and $\eta=0.5$, heat capacity has one zero point at $ r_{+}=r_{1}=0.651 $, which corresponds to a physical limitation point, and also this corresponds to $ T=0 $ ($r_+ = r_m $). Moreover it has two divergence points at $ r_{+}=r_{2}=1.27 $ and $r_{+}=r_{3}=2.76 $, which  demonstrate phase transition critical points of a charged AdS black hole with a global monopole. In other words, the heat capacity is negative at $ r_{+}<r_{1} $, which means that the black hole system is unstable. Then, at $r_{1} <r_{+}<r_{2} $, heat capacity is positive, which means that it is in stable phase. Afterward, at  $r_{2} <r_{+}<r_{3} $, it falls in to a negative region (unstable phase) and, at  $r_{+}>r_{3} $, it becomes stable.
Moreover, from Fig. \ref{pic:C} (c)-(h), by increasing the value of $Q$, $l$ and $\eta$, the  phase transitions occur at different locations.

 Fig.~\ref{pic:P} shows $P-v$ diagrams of the charged AdS black hole with a global monopole, in extended phase space for $ Q = 0.85$, $l=1.5$ and different values of $\eta$.  
Under a critical temperature ($T_{c}$), a critical behavior  appears and with larger values of $ \eta $, this behavior reduces. Also, with larger values of $ \eta $,  $v_c$ increases while $T_{c}$ and $P_{c}$ decrease.

\clearpage

\section{Thermodynamic geometry}\label{sub1}
\begin{figure}[h]
	\centering
	\subfigure[]{
		\includegraphics[width=0.4\textwidth]{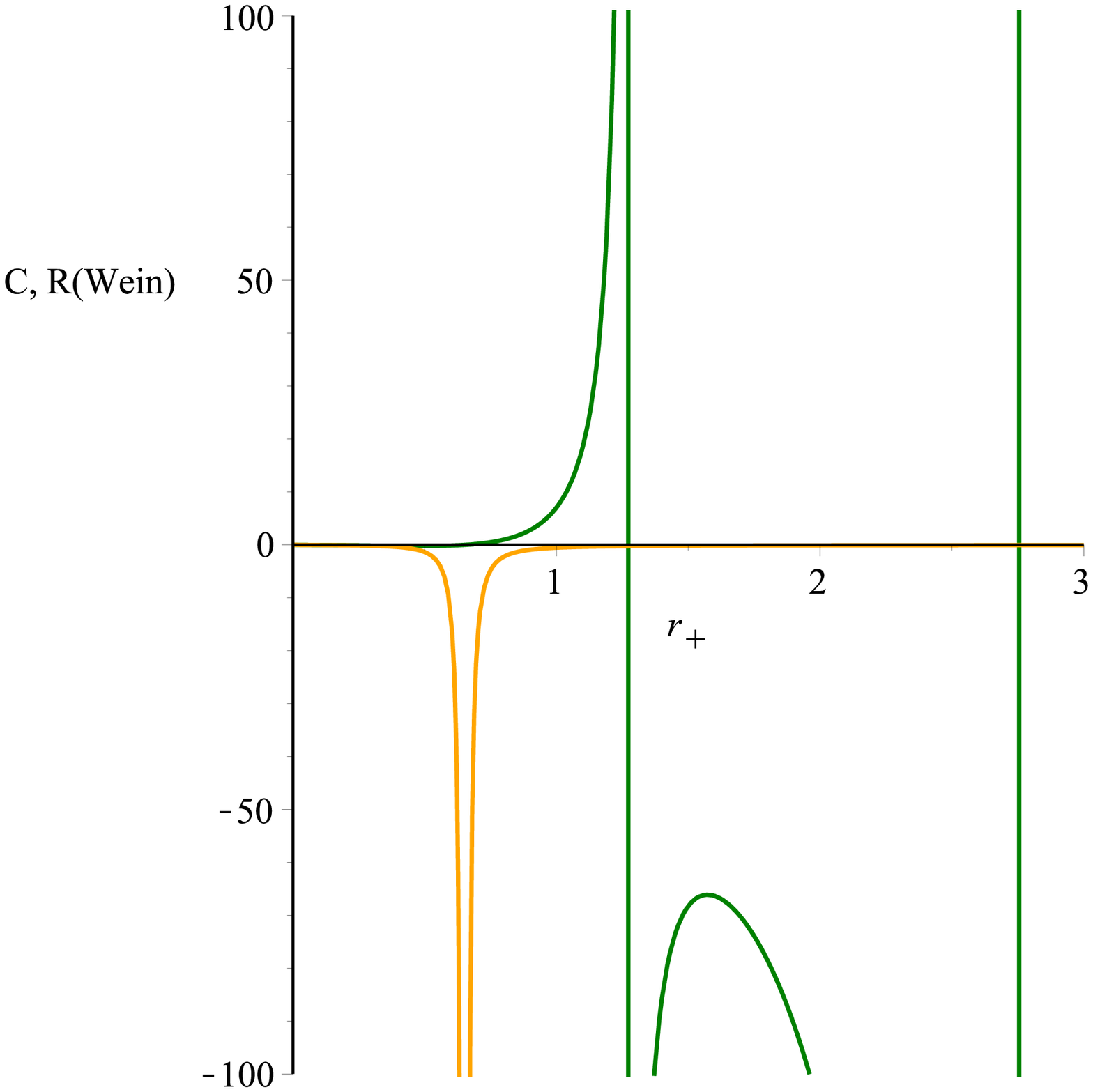}
	}
	\subfigure[Closeup of figure (a)]{
		\includegraphics[width=0.4\textwidth]{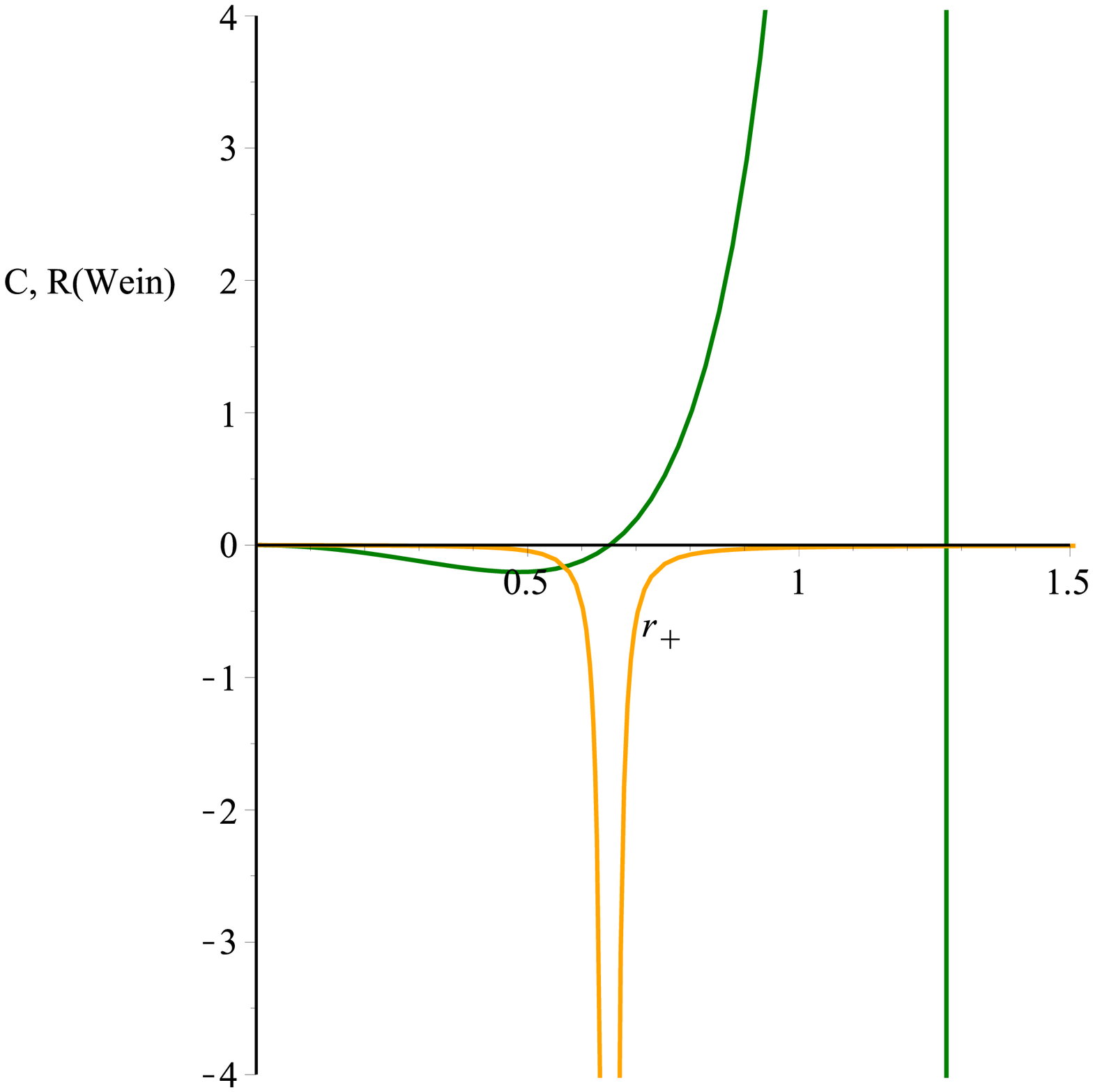}
	}
	\subfigure[]{
		\includegraphics[width=0.4\textwidth]{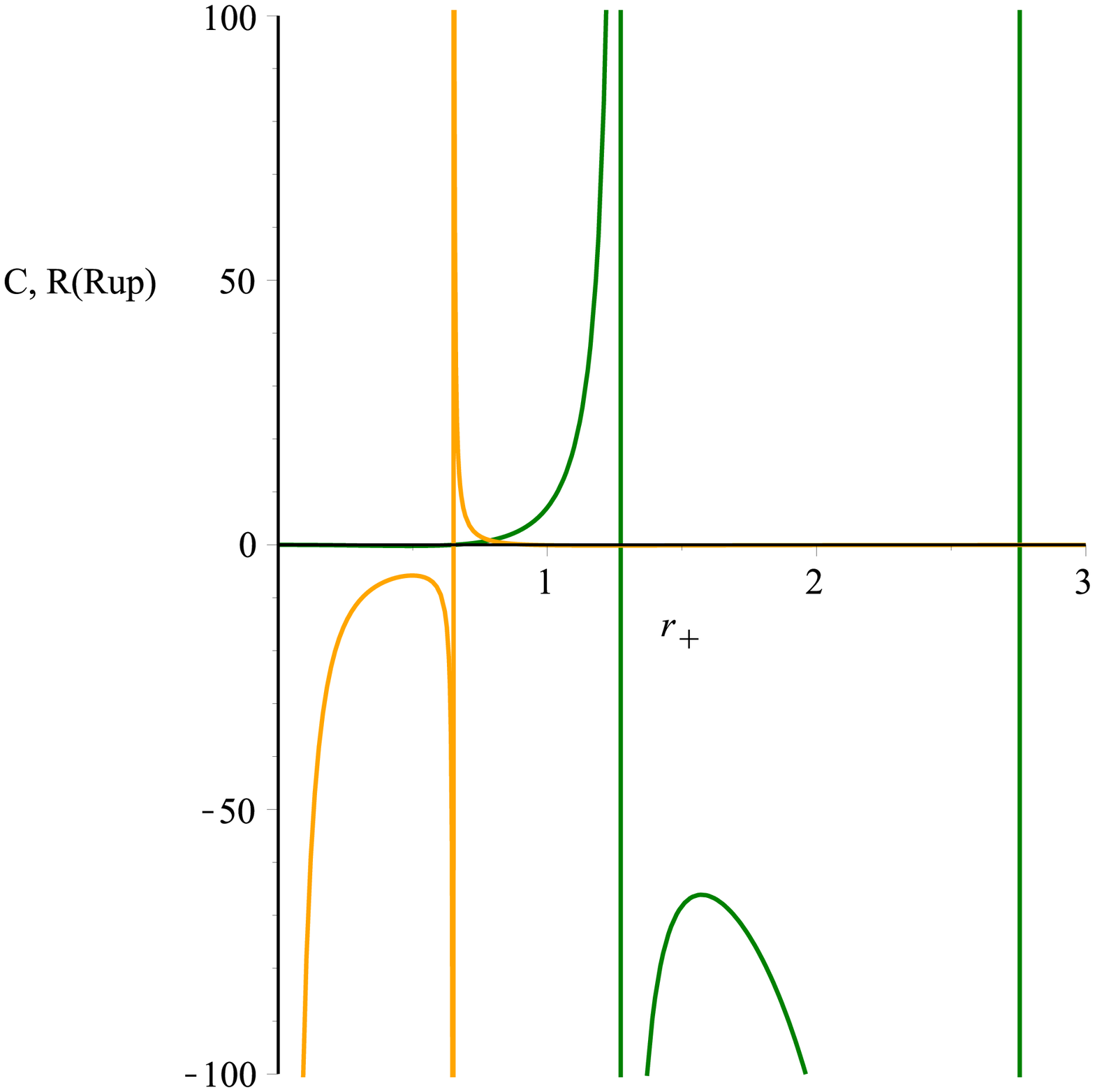}
	}
	\subfigure[Closeup of figure (c)]{
		\includegraphics[width=0.4\textwidth]{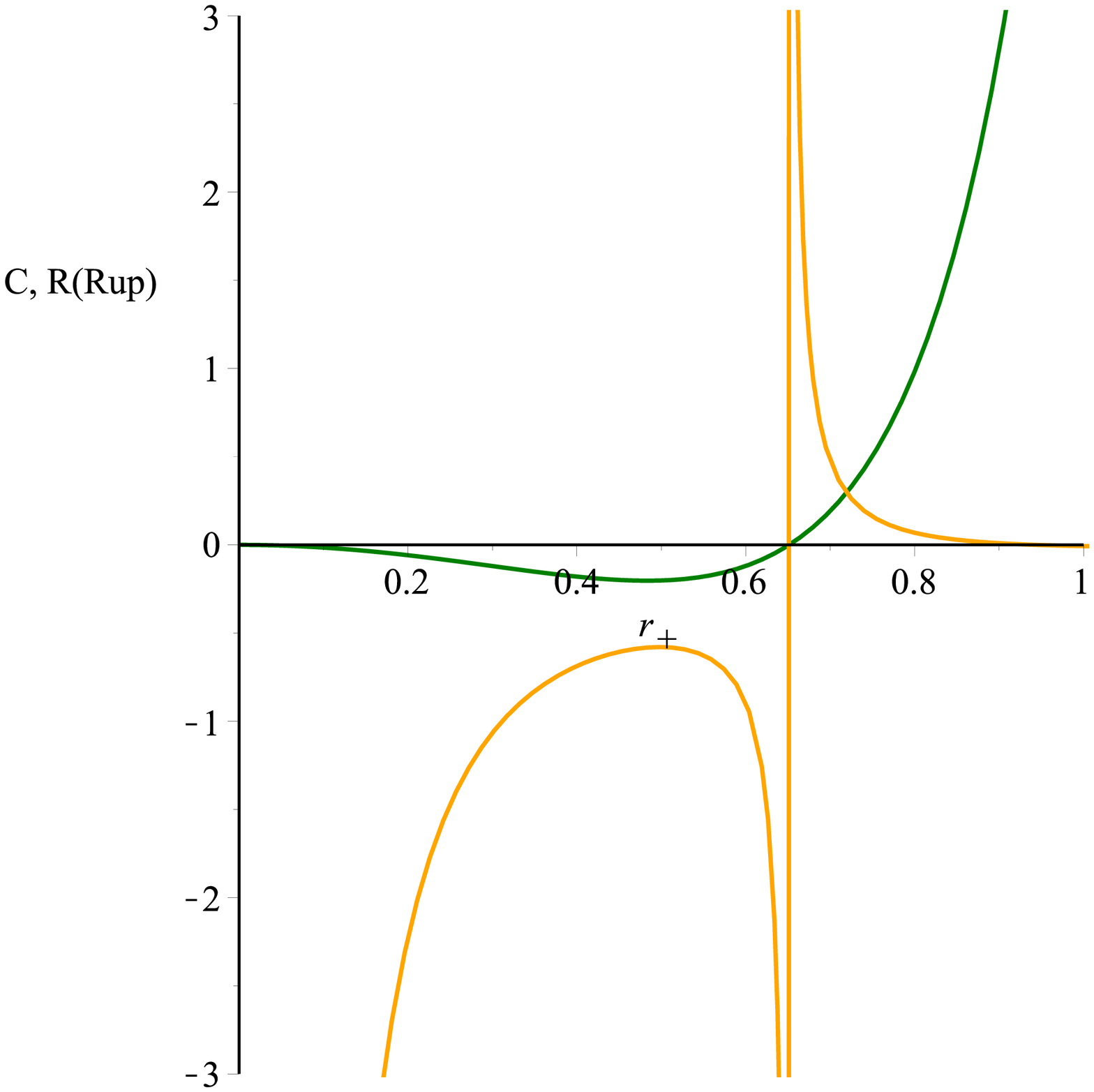}
	}
	\caption{Curvature scalar variation of Weinhold and Ruppeiner metrics (orange line), and also heat capacity variation (Green line), in terms of horizon radius $ r_{+} $ for $Q=0.5$, $l=5.263 $, $\eta=0.5 $.}
	\label{pic:RWeinRup}
\end{figure}

In this section, we study the geometric structure of Weinhold, Ruppiner, Quevedo and HPEM formalisms, and investigate phase transition of a charged AdS black hole with a global monopole. 
The Weinhold geometry is specified in
mass representation as~\cite{Weinhold}
\begin{equation}\label{Weinhold}
	g^{W}_{i j}=\partial _{i}\partial _{j}M(S,l, Q),
\end{equation}
and the line element for a charged AdS black hole with a global monopole is given by \cite{Soroushfar:2016nbu}
\begin{eqnarray}
	ds^{2}_{W}&=&M_{SS}dS^{2}+M_{ll}dl^{2}+M_{QQ}dQ^{2}+ 2M_{Sl}dSdl 
	+ 2M_{SQ}dSdQ +2M_{lQ}dl dQ .
\end{eqnarray}
Therefore the relevant matrix is
\begin{equation}
	g^{W}=\begin{bmatrix}
		M_{S S} & M_{S l} & M_{S Q}\\
		M_{l S} & M_{l l} &0\\
		M_{Q S} & 0& M_{Q Q} 
	\end{bmatrix}.
\end{equation}
So, the curvature scalar 
of the Weinhold metric ($ R^{W} $) is given by 
\begin{equation}
	R^{W}=\frac{l ^{2}~\pi ^{3/2}~S ^{1/2}~\left(\pi ^{2}~Q ^{2}~l ^{2}-\pi ~S ~\eta ^{2}~l ^{2}+\pi ~S ~l ^{2}+9~S ^{2}\right)\left( {1-\eta ^{2}}\right)^{1/2}}{\left(\pi ^{2}~Q ^{2}~l ^{2}+\pi ~S ~\eta ^{2}~l ^{2}-\pi ~S ~l ^{2}-3~S ^{2}\right)^{2}~}.
\end{equation}
The other formalism that we consider here  is the Ruppiner metric. The Ruppiner metric in the thermodynamic system is introduced as~\cite{Ruppeiner,Salamon,Mrugala:1984}
\begin{equation}
	ds^{2}_{R}=\frac{1}{T}ds^{2}_{W},
\end{equation}
and the relevant matrix is 
\begin{equation}
	g^{R}=\left(\frac {4 {l}^{2}{\pi }^{3/2}S \sqrt{-{  {S({\eta}^{2}-1)}{}}} }{{\pi }^{2}{Q}^{2}{l}^{2}+\pi \,S{\eta}^{2}{l}^{2}-\pi \,S{l}^{2}-3\,{S}^{2}} \right)
	\begin{bmatrix}
		M_{S S} & M_{S l} & M_{S Q}\\
		M_{l S} & M_{l l} &0\\
		M_{Q S} & 0& M_{Q Q} 
	\end{bmatrix}.
\end{equation}
Therefore, the curvature scalar of the Ruppiner formalism is calculated by 
\begin{equation}
	R^{Rup}=-\frac {\pi \,{l}^{2} \left( 2\,\pi \,{Q}^{2}+S{\eta}^{2}-S \right) }{ \left( {\pi }^{2}{Q}^{2}{l}^{2}+\pi \,S{\eta}^{2}{l}^{2}-\pi \,S{l}^{2}-3\,{S}^{2} \right) \\
		\mbox{}S}.
\end{equation}
The resulting curvature scalar of Weinhold and Ruppiner metrics  are plotted, in terms of horizon radius $ r_{+} $, to investigate thermodynamic phase transition (see Fig.~\ref{pic:RWeinRup}).

As already noted in Sec.~\ref{sub}, from Fig. \ref{pic:C}, that heat capacity has one zero point at $ r_{+}=0.651 $, which represents a physical limitation point  and   has two divergence points at $ r_{+}=1.27 $ and $r_{+}=2.76 $, which  demonstrate phase transition critical points of a charged AdS black hole with a global monoploe \cite{EslamPanah:2018ums}.
Also, from Fig.~\ref{pic:RWeinRup}, we see that the curvature scalar of Weinhold and Ruppeiner metrics has one singular point at $ r_{+}=0.651 $, which   coincides only with zero point of the heat capacity (physical limitation point).

Now, we use the Quevedo and HPEM metrics to investigate the thermodynamic properties of a charged AdS black hole with a global monopole.
For Quevedo formalism, the general form of the metric is given by
~\cite{Quevedo:2006xk}:
\begin{equation}\label{GTD}
	g=\left(E^{c}\frac{\partial\Phi}{\partial E^{c}}\right)\left(\eta_{ab}\delta^{bc}\frac{\partial^{2}\Phi}{\partial E^{c}\partial E^{d}}dE^{a}dE^{d}\right),
\end{equation}
where
\begin{equation}
	\frac{\partial\Phi}{\partial E^{c}}=\delta_{cb}I^{b}.
\end{equation}
Here, $ \Phi $ is the thermodynamic potential, $ E^{a} $ and $ I^{b} $ are the extensive and intensive 
thermodynamic variables, respectively.
Moreover, the generalized HPEM metric with $ n $ extensive variables ($ n\ge 2 $) is given by \cite{sud,Hendi:2015rja,Hendi:2015xya,EslamPanah:2018ums}
\begin{equation}
	ds_{HPEM}^{2}= \dfrac{SM_{S}}{\left(\prod_{i=2}^n \frac{\partial^{2} M}{\partial \chi_{i}^{2}}\right)^{3}}\left(-M_{SS}dS^{2}+\sum_{i=2}^n \left(\frac{\partial^{2} M}{\partial \chi_{i}^{2}}\right)d\chi_{i}^{2}\right) ,
\end{equation}
in which, $ \chi_{i}(\chi_{i}\neq S) $, $ M_{S}=\frac{\partial M}{\partial S} $ and $ M_{SS} =\frac{\partial^{2} M}{\partial S^{2}} $  are extensive parameters.

\begin{figure}[h]
	\centering
	\subfigure[]{
		\includegraphics[width=0.4\textwidth]{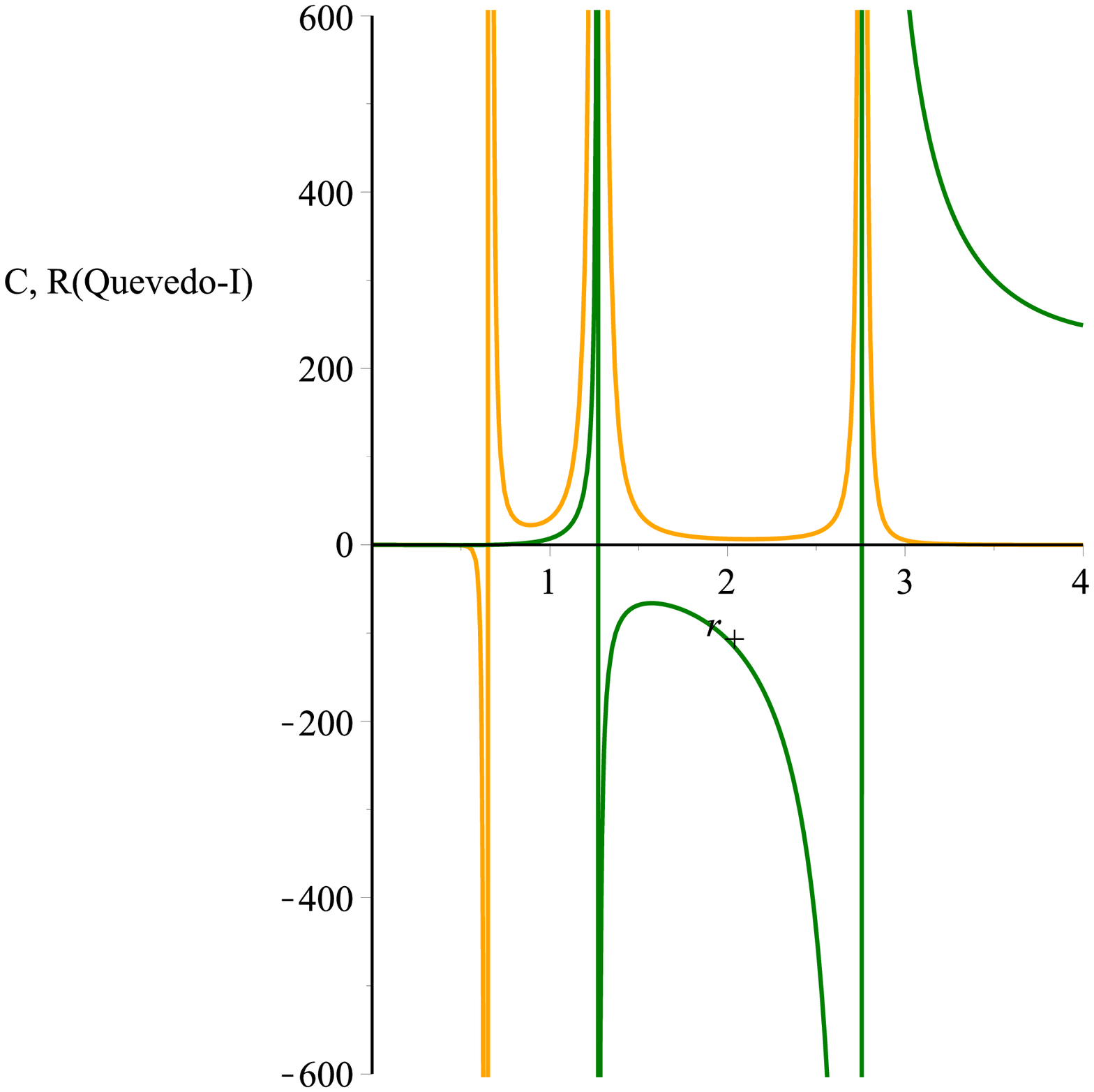}
	}
\subfigure[Closeup of figure (a)]{
	\includegraphics[width=0.4\textwidth]{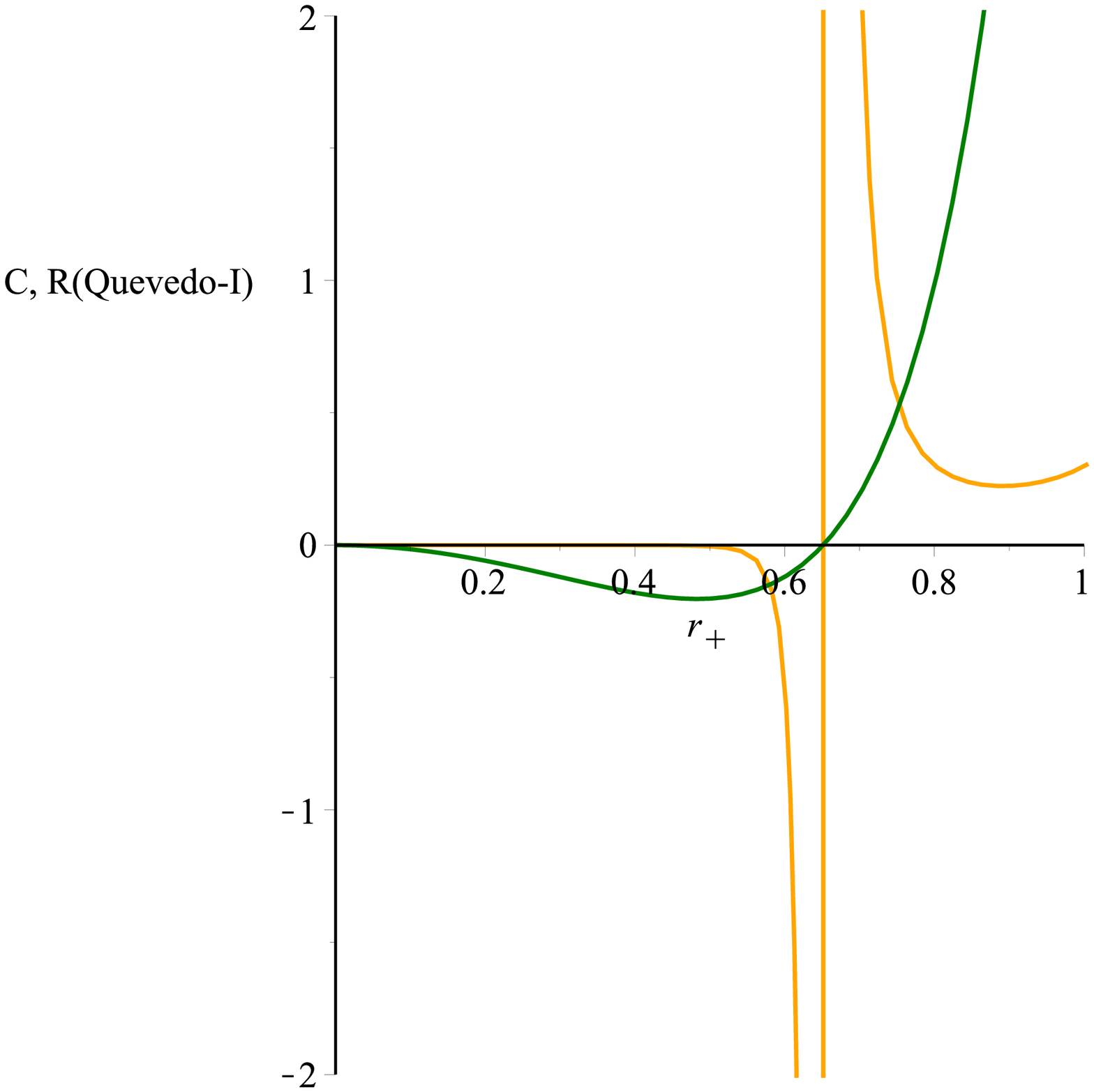}
}
	\subfigure[]{
		\includegraphics[width=0.4\textwidth]{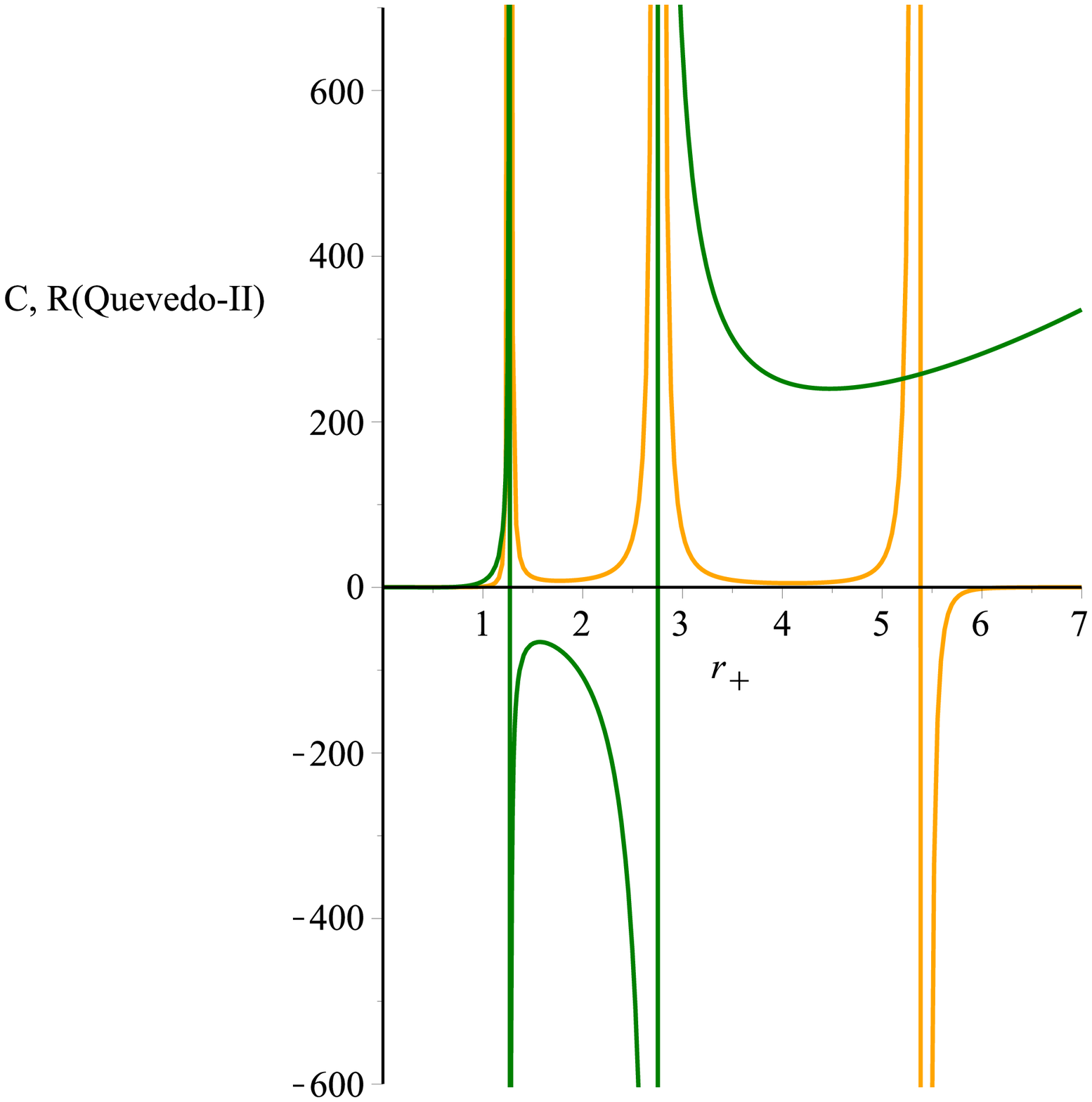}
	}
\subfigure[Closeup of figure (c)]{
	\includegraphics[width=0.4\textwidth]{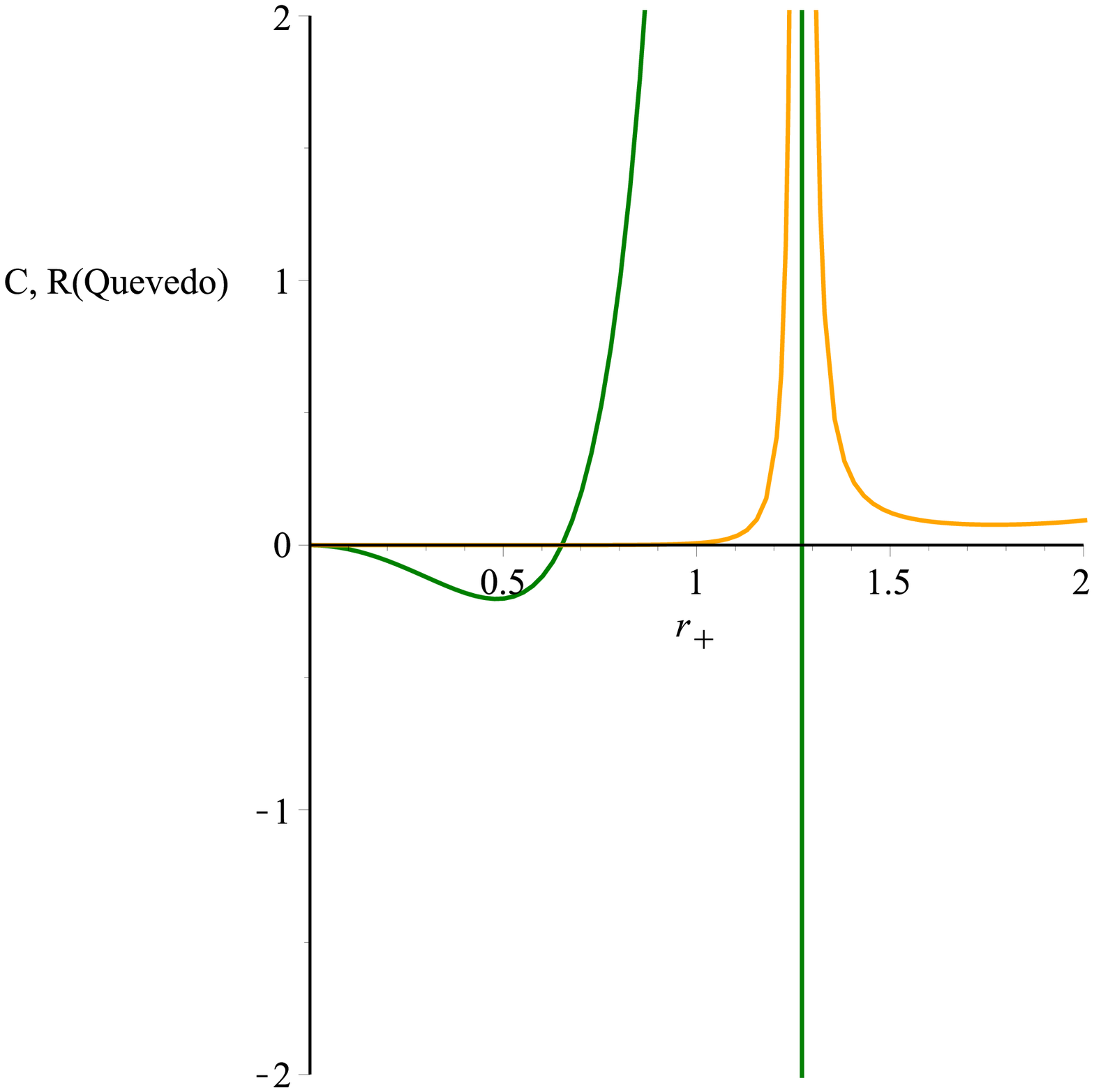}
}
\subfigure[]{
	\includegraphics[width=0.4\textwidth]{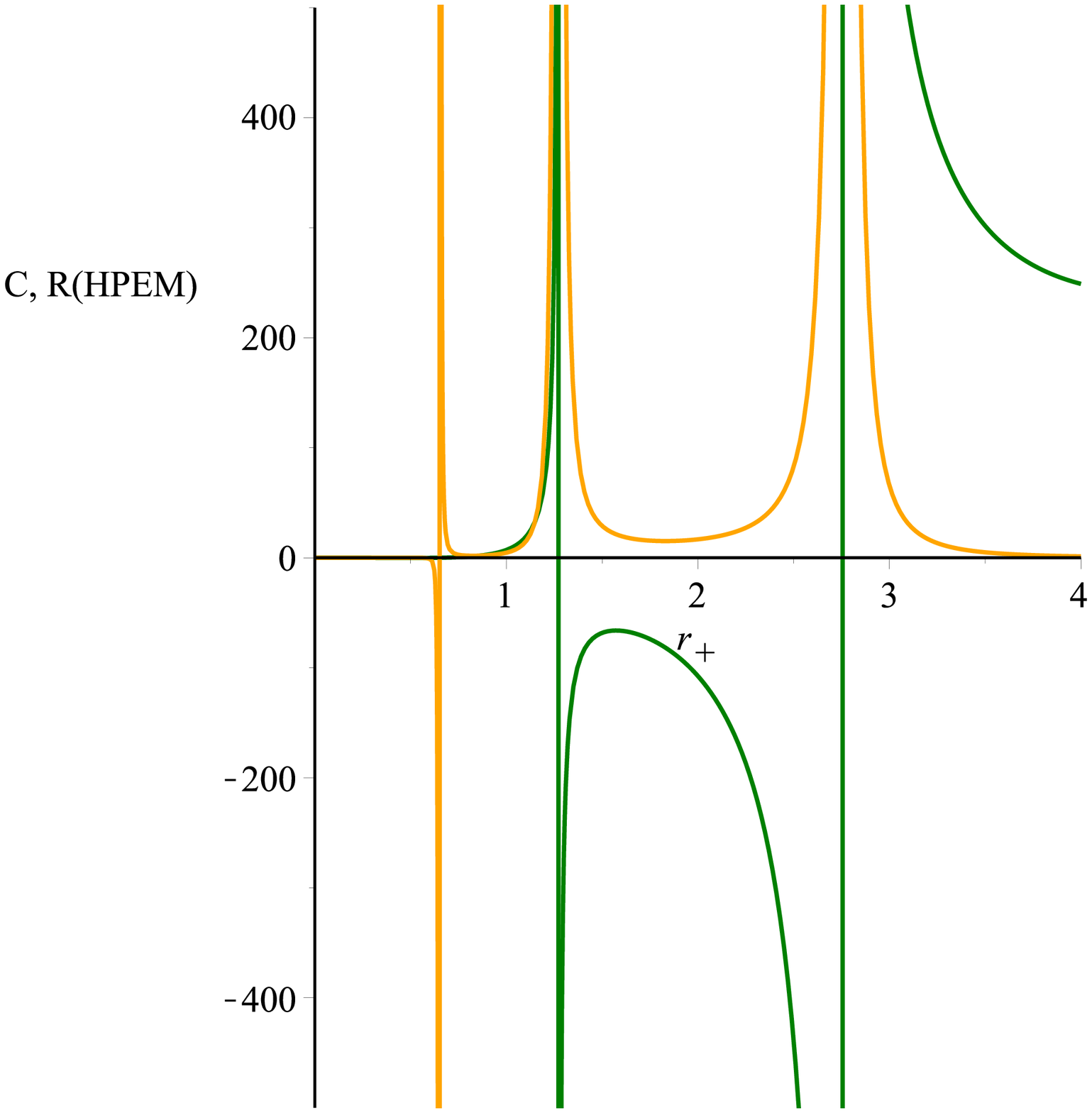}
}
\subfigure[Closeup of figure (e)]{
	\includegraphics[width=0.4\textwidth]{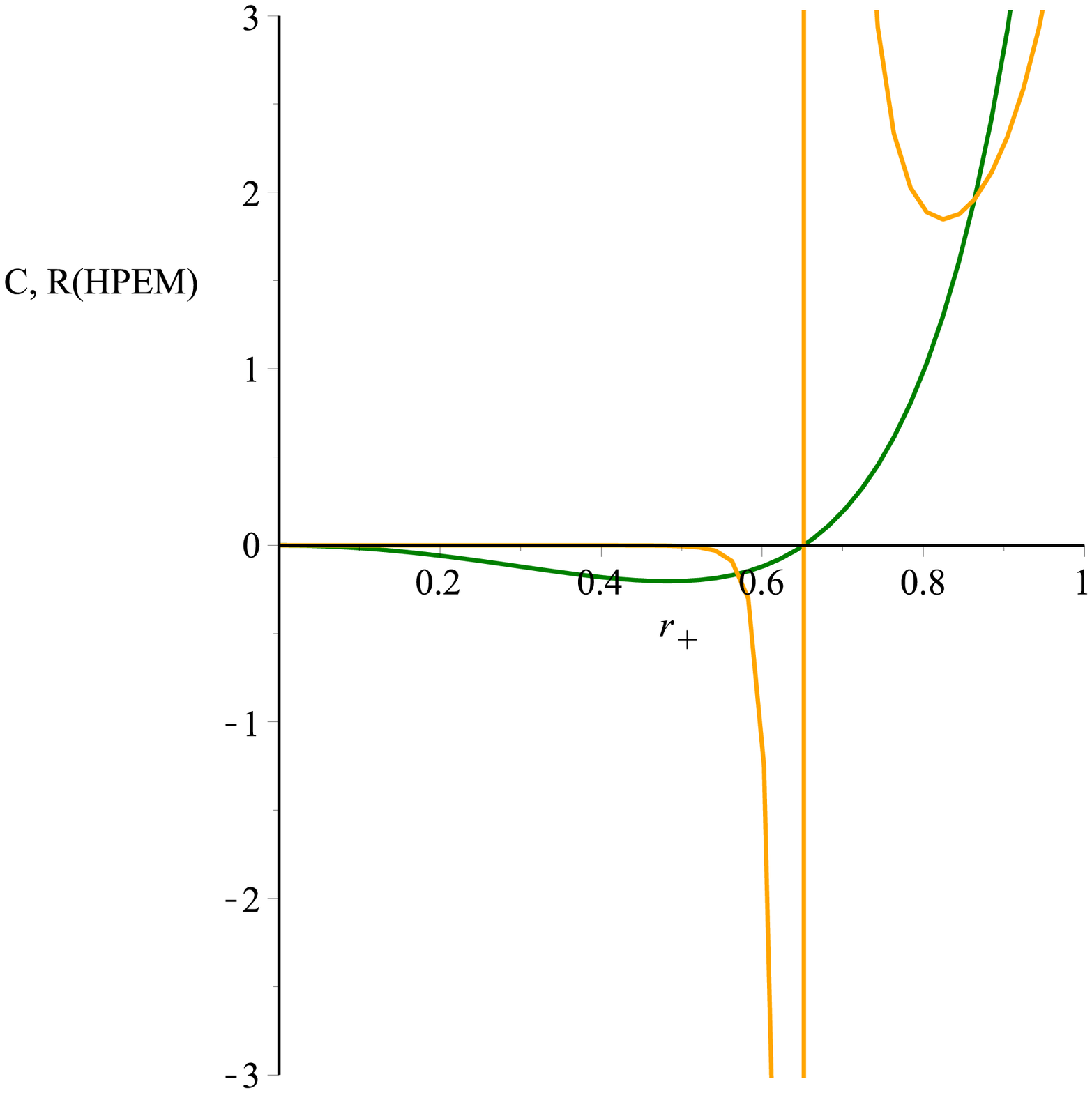}
}
	\caption{Curvature scalar variation of Quevedo and HPEM metrics (orange line) and also heat capacity variation (Green line), in terms of horizon radius $ r_{+} $ for $ Q=0.5 $ and $\eta=0.5 $, $l=5.263 $ .}
	\label{pic:CRQueHPEM}
\end{figure}

The Quevedo and  HPEM metrics can be written, collectively, as \cite{sud,Hendi:2015rja,Hendi:2015xya,EslamPanah:2018ums}
\begin{equation}
	ds^{2}= \begin{cases}
		(SM_{S}+lM_{l}+QM_{Q})(-M_{SS}dS^{2}+M_{ll}dl^{2}+M_{QQ}dQ^{2})   & $Quevedo Case I  $ \\
		
		SM_{S}(-M_{SS}dS^{2}+M_{ll}dl^{2}+M_{QQ}dQ^{2})   &$Quevedo Case II  $    \\
		
		\dfrac{SM_{S}}{\left(\frac{\partial^{2} M}{\partial l^{2}}\frac{\partial^{2} M}{\partial Q^{2}}\right)^{3}}\left(-M_{SS}dS^{2}+M_{ll}dl^{2}+M_{QQ}dQ^{2}\right) & $ HPEM $.
	\end{cases}
\end{equation}
Meanwhile, these metrics have following denominator for their Ricci scalars \cite{sud,Hendi:2015xya,EslamPanah:2018ums}:
\begin{equation}
	\mbox{denom}(R)= \begin{cases}
		2M_{SS}^{2}M_{ll}^{2}M_{QQ}^{2}(SM_{S}+lM_{l}+QM_{Q})^{3}   & $Quevedo Case I  $ \\
		
		2S^{3}M_{SS}^{2}M_{ll}^{2}M_{QQ}^{2}M_{S}^{3}  & $Quevedo Case II  $  \\
		
		2S^{3}M_{SS}^{2}M_{S}^{3} & $ HPEM  $.
	\end{cases}
\end{equation}
The above mentioned equations are solved and plotted in terms of horizon radius $ r_{+} $ (see Fig.~\ref{pic:CRQueHPEM}).

 Moreover, it can be observed from Fig.~\ref{pic:CRQueHPEM}(a),(b) that the curvature scalar of Quevedo (case-I) metric has three singular points at $ r_{+}=0.651 $, $ r_{+}=1.27 $ and $r_{+}=2.76 $, which coincide with zero point (physical limitation point) and divergence points (transition critical points) of heat capacity, respectively. 

In addition, Fig.~\ref{pic:CRQueHPEM}(c),(d) shows that the curvature scalar of Quevedo (case-II) metric has three singular points at $ r_{+}=1.27 $, $r_{+}=2.76 $ and $r_{+}=5.39 $, in which two of them only coincide with divergences points (transition critical points) of heat capacity and  no singular point  coincides with zero point of heat capacity.

Furthermore, in case of HPEM metric, as shown in Fig.~\ref{pic:CRQueHPEM}(e),(f), the divergence points of the Ricci scalar (i.e. $ r_{+}=0.651 $, $ r_{+}=1.27 $ and $r_{+}=2.76 $) are coincident with zero point (physical limitation point) and divergence points (transition critical points) of heat capacity, respectively. So, the divergence points of the Ricci scalar of HPEM and Quevedo (case-I) metrics coincide with both types of phase transitions of the heat capacity. Therefore, it can be claimed that one can extract more information from HPEM and Quevedo cases as compared with Weinhold and Ruppeiner metrics. 

 \clearpage
\section{Conclusions}\label{section4}
In order to discuss phase transition of a charged AdS black hole with a
global monopole through geometrical thermodynamics, we have first written metric for a charged AdS black hole with global monopole. We have shed light on
  the Hawking temperature, specific heat and electric potential for the system. Here, we have 
derived critical parameters also for this black hole.
We have studied the behavior of temperature in terms of horizon radius for different values of charge, AdS radius and monopole parameter and observed that Hawking temperature 
first increases to a maximum point and then starts falling  as  horizon radius increases.  Finally, after reaching a certain value of  event horizon radius, the Hawking temperature  only increases  with horizon radius.  

Moreover, we have plotted  heat capacity with respect to horizon radius for different values of charge, AdS radius and monopole parameter. Interestingly, we have found that  for certain values of these parameters, heat capacity has one zero point describing a physical limitation point. Also,  heat capacity has two divergence points, which  describe phase transition critical points of a charged AdS black hole with a global monopole. 
We have plotted pressure-volume curve also for the charged AdS black hole with a global monopole for  the specific values of the parameters.  
Below critical temperature, a critical behavior  has been observed  which changes  with larger values of energy scale. Remarkably, for larger values of the energy scale (monopole parameter), in contrast to  critical temperature and critical pressure, only critical volume increases.

Subsequently, we have provided the geometric structure of Weinhold, Ruppiner, Quevedo and HPEM
formalisms in order to investigate  phase transition of a charged AdS black hole with a global monopole. In this regard, we have computed first  the curvature scalar of Weinhold and  
Ruppeiner metrics. We have plotted these scalar curvature with horizon radius. These plots suggested that   the curvature scalar of Weinhold and Ruppeiner metrics has one singular point, which   coincides only with zero point of the heat capacity. 

The Quevedo and HPEM formalisms are also implemented  to investigate the thermodynamic properties of a charged AdS black hole with a global monopole.
We have observed that  the curvature scalar of Quevedo  metric (case-I) has three singular points, which coincide with zero point (physical limitation point) and divergence points (transition critical points) of heat capacity, respectively. However, in case of curvature scalar of Quevedo (case-II) metric, there exist three singular points; two of them only coincide with divergences points (transition critical points) of heat capacity and no point coincides with zero point of heat capacity.
In case of HPEM metric, The curvature scalar exhibits divergence points which are coincident with zero point (physical limitation point) and divergence points (transition critical points) of heat capacity, respectively. Here, we concluded that the divergence points of the Ricci scalar of HPEM and Quevedo (case-I) metrics coincide with both types of phase transitions of the heat capacity.  These analysis suggest that one can get more information about the phase transition for the   charged AdS black hole with a
global monopole from HPEM and Quevedo methods in comparison to the Weinhold and Ruppeiner cases.

\end{document}